\documentclass[11pt]{article}
\usepackage[margin=0.7in]{geometry}
\usepackage{amsmath,amssymb}
\usepackage{graphicx}
\usepackage{cite}
\usepackage{hyperref}

\usepackage{graphicx}

\renewenvironment{abstract}
	{\quotation}
	{\endquotation}
    
\usepackage[utf8]{inputenc}
\DeclareUnicodeCharacter{2032}{\ensuremath{\prime}}
\date{}

\makeatletter
\renewcommand{\fnum@figure}{\textbf{Figure \thefigure}}
\renewcommand{\fnum@table}{\textbf{Table \thetable}}
\makeatother

\usepackage{url}

\def\scititle{
	A Quantum Framework for Protein Binding-Site Structure Prediction on Utility-Level Quantum Processors
}
\title{\bfseries \boldmath \scititle}

\author{
Yuqi Zhang$^{1}$, Yuxin Yang$^{2}$, William Martin$^{2}$, Kingsten Lin$^{2}$, Zixu Wang$^{1}$,\\
Cheng-Chang Lu$^{3}$, Weiwen Jiang$^{4}$, Ruth Nussinov$^{5,6}$, Joseph Loscalzo$^{7}$,\\
Qiang Guan$^{1\ast}$, Feixiong Cheng$^{2,8,9,10\ast}$\\
\\
\normalsize{$^{1}$Department of Computer Science, Kent State University, Kent, OH, USA}\\
\normalsize{$^{2}$Cleveland Clinic Genome Center, Lerner Research Institute, Cleveland Clinic,}\\ \normalsize{Cleveland, OH, USA}\\
\normalsize{$^{3}$Qradle Inc., Kent, OH, USA}\\
\normalsize{$^{4}$Department of Electrical and Computer Engineering, George Mason University, Fairfax, VA, USA}\\
\normalsize{$^{5}$Computational Structural Biology Section,  National Cancer Institute, }\\
\normalsize{Frederick National Laboratory for Cancer Research in the Cancer Innovation Laboratory,}\\
\normalsize{Frederick, MD, USA}\\
\normalsize{$^{6}$Department of Human Molecular Genetics and Biochemistry,}\\ 
\normalsize{Sackler School of Medicine, Tel Aviv University, Tel Aviv, Israel}\\
\normalsize{$^{7}$Department of Medicine, Brigham and Women's Hospital, }\\
\normalsize{Harvard Medical School, Boston, MA, USA}\\
\normalsize{$^{8}$Genomic Medicine Institute, Lerner Research Institute, Cleveland Clinic, Cleveland, OH, USA}\\
\normalsize{$^{9}$Department of Molecular Medicine, Cleveland Clinic Lerner College of Medicine,}\\ 
\normalsize{Case Western Reserve University, Cleveland, OH, USA}\\
\normalsize{$^{10}$Case Comprehensive Cancer Center, Case Western Reserve University,}\\
\normalsize{School of Medicine, Cleveland, OH, USA}\\
\\
\normalsize{$^\ast$To whom correspondence should be addressed; E-mail: qguan@kent.edu; chengf@ccf.org.}
}

\begin{document}

\maketitle

\begin{abstract} \bfseries \boldmath
Accurate prediction of protein active site structures remains a central challenge in structural biology, particularly for short and flexible peptide fragments where conventional and simulation-based methods often fail. Here, we present a quantum computing framework specifically developed for utility-level quantum processors to address this problem. Starting from an amino acid sequence, we formulate the structure prediction task as a ground-state energy minimization problem using the Variational Quantum Eigensolver (VQE). Amino acid connectivity is encoded on a tetrahedral lattice model, and structural constraints—including steric, geometric, and chirality terms—are mapped into a problem-specific Hamiltonian expressed as sparse Pauli operators. The optimization is executed via a two-stage architecture separating energy estimation and measurement decoding, allowing noise mitigation under realistic quantum device conditions. We evaluate the framework on 23 randomly selected real protein fragments from the PDBbind dataset, as well as 7 fragments from therapeutically relevant proteins, and run the experiments on the IBM–Cleveland Clinic quantum processor. Structural predictions are benchmarked against both AlphaFold3 (AF3) and classical simulation–based approaches using identical postprocessing and docking procedures. Our quantum method outperformed both AF3 and classical models in RMSD (Root-Mean-Square Deviation) and docking efficacy. This work demonstrates, for the first time, a complete end-to-end pipeline for biologically relevant structure prediction on real quantum hardware, highlighting its engineering feasibility and practical advantage over existing classical and deep learning approaches.

\end{abstract}

\section*{Introduction}
\noindent
Predicting the local three-dimensional structures of proteins, particularly at ligand-binding sites, is a central challenge in structural biology with broad implications for molecular function annotation, drug design, and protein engineering~\cite{liang1998anatomy}. The conformational landscape of these regions is often characterized by high flexibility and functional specificity, making accurate modeling both scientifically critical and technically challenging~\cite{perot2010druggable}. Traditional physics-based methods, such as molecular dynamics (MD), can in principle sample native-like structures but suffer from prohibitive computational costs and scalability issues~\cite{hansmann1999new}. By contrast, recent breakthroughs in deep learning, exemplified by AlphaFold2 and AlphaFold3, have achieved remarkable success in full-length protein structure prediction~\cite{jumper2021highly,abramson2024accurate}. However, their performance substantially deteriorates when applied to short peptide fragments or highly flexible local domains, where limited sequence context and sparse evolutionary information lead to structural ambiguities and reduced accuracy~\cite{scardino2023good,chakravarty2025proteins}. In particular, short fragments that constitute ligand-binding pockets typically contain fewer than 20 amino acids and exhibit complex energetic constraints that are difficult to capture with purely data-driven models.

While several approaches have explored quantum computing for protein structure prediction—including formulations as binary optimization tasks, quantum annealing schemes, and the Quantum Approximate Optimization Algorithm (QAOA)~\cite{romero2025protein}—their validation in realistic, biologically relevant settings remains limited. Most prior methods have been evaluated only on toy examples or simulations and often lack the accuracy needed for downstream applications such as ligand docking. This gap underscores the need for frameworks that can operate under limited input information (e.g., short fragments), make efficient use of quantum resources, and deliver predictions of sufficient precision to support molecular docking and design. Quantum computing offers a fundamentally new paradigm for tackling such problems by exploiting quantum parallelism, entanglement, and interference. Although current hardware is constrained by noise, decoherence, and limited qubit counts—commonly referred to as utility-level processors—it has reached a level of maturity that enables early-stage, domain-specific applications. Among hybrid quantum-classical algorithms, the Variational Quantum Eigensolver (VQE) is particularly well-suited for energy landscape optimization~\cite{papanikolaou2025hybrid,wei2020full}. By encoding physical constraints into parameterized quantum circuits, VQE enables estimation of ground-state energies for Hamiltonians constructed from simplified structural models.

In this work, we present an end-to-end quantum computing framework for predicting the local structure of protein binding-site fragments on utility-level quantum hardware. Our approach begins with an amino acid sequence and maps its spatial configuration to a tetrahedral lattice, where each residue is represented as a node with four directional connectivity options. These configurations are encoded as quantum measurement states corresponding to discrete Hamiltonians that incorporate steric constraints, chirality terms, geometric restrictions, and amino acid interactions. To mitigate the effects of hardware noise, we design a two-stage execution architecture that separates energy estimation from structural decoding. This strategy enhances execution stability while preserving fidelity in structural recovery. We validate our framework using short peptide sequences extracted from the PDBbind database~\cite{wang2004pdbbind,wang2005pdbbind,liu2015pdb} and execute them on the IBM–Cleveland Clinic 127-qubit superconducting quantum processor. For benchmarking, we compare our results against both AlphaFold3 and classical simulation–based prediction methods, followed by uniform postprocessing and docking evaluation using AutoDock~\cite{huey2012using,gaillard2018evaluation,trott2010autodock,eberhardt2021autodock}. Across all test fragments, the quantum framework consistently achieves lower RMSD values and more favorable docking affinities than both AlphaFold3 and classical models, thereby demonstrating the feasibility and practical advantage of quantum-assisted structural prediction on real devices.

The key contributions of this work are as follows: 
(1) We introduce a coarse-grained, vector-based lattice encoding and corresponding Hamiltonian formulation tailored for short peptide fragments at binding pockets, which reduces side-chain complexity and produces a smoother, noise-resilient energy landscape.
(2) We develop an end-to-end quantum computing framework that separates energy estimation from structural decoding, thereby enhancing robustness against hardware noise.  
(3) We demonstrate, for the first time, biologically relevant structure prediction for docking-oriented fragments on a real 127-qubit superconducting quantum processor, achieving superior RMSD and docking performance compared to both AlphaFold3 and classical simulation–based approaches. 
Together, these advances highlight a practical role for today’s noisy quantum hardware in structural biology and establish a blueprint for integrating quantum computing into protein docking and design workflows.

\section*{Results}

\subsection*{A Quantum Engineering Framework for Structure Prediction}

\begin{figure*}
    \centering
    \includegraphics[width=1\linewidth]{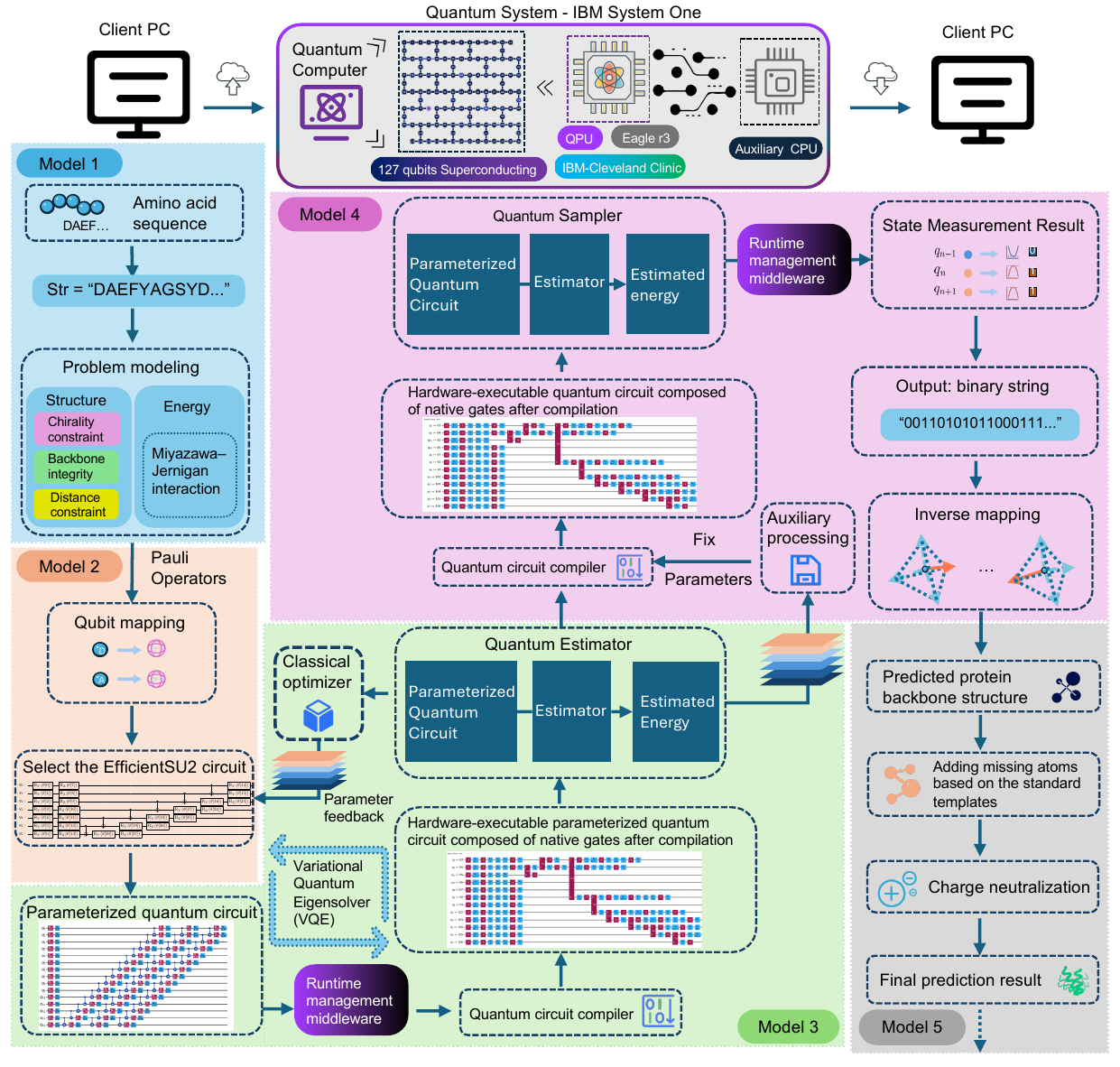}
    \caption{\textbf{End-to-End Hardware-Executable Quantum Workflow for Protein Structure Prediction.} The framework takes an amino acid sequence as input, encodes structural constraints into Pauli operators, and constructs a parameterized quantum circuit using a selected ansatz. A hybrid VQE is executed on IBM Quantum hardware via Qiskit Runtime to optimize circuit parameters. The lowest-energy binary outcome is reverse-mapped to a predicted backbone structure, followed by postprocessing for downstream applications. The framework consists of five submodules: \textbf{(1)} Problem Modeling Module: Models the input sequence as Pauli operators and constructs the corresponding Hamiltonian. \textbf{(2)} Quantum Circuit Construction Module: Selects the circuit architecture, builds a parameterized quantum circuit, and associates it with the Hamiltonian. \textbf{(3)} Hybrid Variational Optimization Module: Optimizes the parameters of the quantum circuit using a quantum-classical hybrid algorithm (VQE) to search for low-energy states. \textbf{(4)} Measurement Module: Reconstructs a fixed-parameter quantum circuit using the optimized parameters, performs measurements, and decodes the results. \textbf{(5)} Postprocessing Module: Processes the predicted outputs to generate complete protein structures suitable for downstream biological tasks such as docking or simulation.}
    \label{fig1}
\end{figure*}

To enable the prediction of protein binding site fragment structures on real quantum hardware, we developed a fully deployable, end-to-end quantum computing framework tailored for utility-level quantum processors (Figure~\ref{fig1}). This framework takes amino acid sequences as input and outputs three-dimensional conformations that satisfy physical and geometric constraints. The system is designed with engineering completeness, modular extensibility, and hardware compatibility in mind, supporting downstream structural and functional evaluations.

The framework consists of the following core stages: 
(1)~Sequence-based quantum modeling, where the input amino acid sequence is used to construct a quantum formulation of the structural problem, incorporating Miyazawa–Jernigan interactions and additional geometric constraints, which are encoded as Pauli operators. (2)~Parameterized circuit construction, where the number of required qubits is determined and a parameterized quantum circuit is built using a selected ansatz, with random initialization of parameters and Hamiltonian association. (3)~Hybrid variational optimization, in which the Qiskit Runtime middleware orchestrates VQE execution across quantum and classical resources, iteratively updating circuit parameters to minimize system energy on real quantum hardware. (4)~Measurement and structural decoding: A new fixed-parameter quantum circuit is constructed by combining the optimized parameters with the same ansatz, followed by a second-stage compilation to produce a hardware-executable measurement circuit. The circuit is then executed with a fixed number of shots as specified by the user-defined precision. The resulting bitstrings are statistically processed, and the most probable output is reverse-mapped into a spatial backbone vector. And (5)~Postprocessing for downstream usability, including atom completion and charge neutralization to generate biologically usable protein structures for tasks such as docking or simulation. As illustrated in Figure~\ref{fig1}, the framework takes an amino acid sequence represented as a string as input. This sequence is first modeled using the Miyazawa–Jernigan interaction~\cite{miyazawa1999self} along with additional structural constraints. The modeled problem is then translated into the quantum domain in the form of Pauli operators. These Pauli operators are used to determine the required number of qubits and to construct the Hamiltonian of the system. Based on the Pauli formulation, the algorithm automatically constructs a parameterized quantum circuit using a selected ansatz and initializes the variational parameters randomly. The execution is managed through a Qiskit Runtime session~\cite{qiskit2024}, which handles job orchestration. The parameterized circuit and Hamiltonian are uploaded to the quantum processor, where a second-stage compilation is performed. A cross-platform hybrid Variational Quantum Eigensolver (VQE) is then executed, combining quantum sampling with a classical optimizer running locally. The optimization loop in the VQE continues iterating until the minimum energy is found or a predefined termination condition is satisfied. At convergence, the optimal set(s) of variational parameters are bound to the predefined ansatz, and the circuit is recompiled. The finalized quantum circuit is then executed multiple times according to a user-defined precision to perform quantum state measurements. Upon completion of the measurement task, results are retrieved. The measured outcomes are statistically processed on the classical backend, and the highest-probability result—represented as a bitstring (binary string)—is selected. This bitstring is then reverse-mapped using the same encoding rules defined during problem modeling, reconstructing the predicted protein backbone structure. A final postprocessing stage is performed locally, including structural refinement steps such as atom completion and charge neutralization, in preparation for downstream tasks such as molecular docking.

Each residue in the input sequence is mapped to a node on a tetrahedral grid, where four directional edges represent possible backbone extensions. The spatial connectivity between adjacent residues is encoded into segments of a quantum bitstring, which defines the geometric path of the chain. To model physical validity, we construct a sparse Pauli Hamiltonian incorporating chirality constraints, steric repulsion, geometric feasibility, and amino acid interaction energies. This Hamiltonian is optimized using VQE to obtain a quantum state corresponding to the lowest-energy configuration. Given the constraints of current utility-level hardware—including short coherence times, measurement noise, and gate infidelity—we introduce a two-phase execution strategy. In the first phase, the parameterized quantum circuit is used for energy estimation and variational optimization. In the second phase, the optimized parameters from the first phase are fixed and compiled into a stable circuit composed entirely of native quantum gates supported by the processor. This fixed circuit is then measured repeatedly to obtain bitstrings at a user-specified precision. The decoupling of optimization and measurement enhances robustness to noise, increases execution stability, and ensures the reproducibility of structural predictions. The final output of the framework is a set of spatially consistent three-dimensional structures that can be evaluated through structural analysis and ligand docking.

\subsection*{Execution on Utility-Level Quantum Hardware}

To validate the executability and engineering robustness of the proposed framework on real quantum hardware, we deployed the full structure prediction pipeline on the IBM-Cleveland Clinic’s 127-qubit superconducting quantum processor. All tasks were executed directly on physical hardware rather than in simulation environments. Depending on the fragment length and the complexity of encoded amino acid interactions, the number of active qubits required for each task ranged from several dozen to over a hundred. We used a tetrahedral lattice model to discretize the amino acid sequence and constructed a sparse Pauli Hamiltonian incorporating steric repulsion, chirality constraints, geometric feasibility, and residue-level interaction energies. During the variational optimization phase, we employed a parameterized quantum circuit composed of native single-qubit rotation gates (e.g., $\text{Ry}(\theta)$) and controlled-NOT (CX) gates. These circuits were compiled and topology-mapped using Qiskit’s transpiler to ensure compatibility with the IBM-Cleveland Clinic quantum processor. To improve tolerance to hardware noise and maintain structural prediction stability, we adopted a two-stage execution strategy, architecturally separated into a variational optimization module and a dedicated measurement module.  In the first stage, the Variational Quantum Eigensolver (VQE) was used to minimize the expectation value of the constructed Hamiltonian. Each energy evaluation was performed using 2,000 shots for quantum energy estimation, combined with at least 200 iterations of the COBYLA (Constrained Optimization BY Linear Approximations) classical optimizer. In the second stage, the optimized parameter set from the first phase was fixed and used to reconstruct a non-parametric circuit, which was then recompiled into a stable quantum circuit composed entirely of native gates. This fixed circuit was executed 20,000 times to obtain measurement bitstrings representing structural states. This two-phase process ensured reproducibility and robustness under noise-constrained physical environments. For testing, we selected 23 representative peptide fragments from the PDBbind database~\cite{liu2015pdb}. Each fragment originated from a distinct protein-ligand complex with experimentally determined receptor structures. These fragments were chosen to be structurally independent and diverse in both residue composition and sequence ordering, enabling us to assess the generalization capability of our framework across varying input conditions. Table~\ref{tab:execution} summarizes the selected samples’ PDB IDs, residue sequences, sequence lengths, and execution parameters including number of qubits, circuit depth, execution time, and the system’s minimum, maximum, and energy range during optimization. Due to the high computational cost of quantum hardware, we adopted a strategy of low-iteration VQE optimization coupled with high-precision measurements. If resource constraints were relaxed, it would be feasible to achieve better convergence paths and structure prediction accuracy by increasing the number of optimization steps and adopting gradient-based optimizers. The reason why the execution times shown in the table do not increase proportionally with the length of the protein fragments is mainly due to the current task scheduling mechanism on quantum computers. Since the scheduling strategy is not yet fully intelligent, some tasks may be preempted by higher-priority jobs—such as those submitted by accounts with higher access levels or with shorter estimated runtimes—which results in additional session delays and idle time.

The 23 test cases ranged from short peptides with 5 amino acids to medium-length chains of 10–14 residues, ensuring the framework’s compatibility across varying sequence lengths. Several clear trends were observed from the collected execution data: first, the number of qubits increased significantly with sequence length, as more ancillary qubits are required to represent growing residue-residue interactions and constraints. Second, the system energy exhibited a stepwise increase as sequence length increased, with sharp jumps observed at certain threshold lengths. Using the system's energy range as an indicator, we found that increasing the sequence length from 6 to 7 residues led to a jump in energy range from approximately 260 a.u. (atomic units, Hartree energy) to over 600 a.u., an increase of about 131\%. From 7 to 9 residues, energy range increased from around 610 a.u. to approximately 1,350 a.u. (+121\%), and further to over 2,100 a.u. at 10 residues (+56\%). Between 10 and 11 residues, energy range continued to rise to 2,500–2,700 a.u. (+20–30\%), then reached 4,600–5,300 a.u. at 13 residues (+85\%), and finally surpassed 9,000 a.u. at 14 residues (+72\% over 13 residues and +2,700\% over the initial 6-residue case). Both the minimum and maximum energy values also expanded rapidly with sequence length. For example, minimum energy increased from approximately 10 a.u. (5 residues) to over 23,000 a.u. (14 residues), while maximum energy increased from ~15 a.u. to over 32,000 a.u.—representing increases of approximately 230,000\% and 210,000\%, respectively. These nonlinear surges demonstrate the exponential growth of the quantum energy landscape as conformational dimensionality increases. This nonlinear energy amplification reveals two key insights. First, the quantum system is capable of encoding subtle structural variations with high precision: even the addition or removal of a single amino acid causes energy changes distinguishable by the quantum model. Second, the energy of the quantum system, as encoded under our Hamiltonian mapping, is highly correlated with the underlying conformational energy of the protein fragment. Thus, the lowest-energy quantum state produced by VQE directly corresponds to the predicted structure, validating the soundness of our prediction strategy. It is precisely due to this energy amplification behavior that we modularized the measurement stage, allowing it to operate independently with non-parametric circuits for high-precision bitstring sampling. This design preserves prediction fidelity while maintaining engineering tractability under realistic hardware constraints. It should be noted that the observed ``energy amplification'' trend primarily originates from the Hamiltonian construction itself, where the number and magnitude of penalty and interaction terms scale rapidly with sequence length. Quantum computation does not artificially amplify energies, but rather provides a mechanism to faithfully encode and optimize this highly nonlinear energy landscape. In this sense, the amplification reflects the exponential growth of conformational complexity under our Hamiltonian mapping, while the quantum framework enables tractable energy minimization and structural recovery under these conditions.

\begin{table}
\centering
\scriptsize
\caption{\textbf{Summary of quantum structure prediction tasks on IBM-Cleveland Clinic.} Each row corresponds to a predicted protein fragment with its amino acid sequence, number of qubits, energy profile, execution time, and circuit depth.}
\label{tab:execution}
\vspace{0.5em}
\begin{tabular}{l l c c c r r r r c}
\hline
\textbf{PDB ID} & \textbf{Sequence} & \textbf{Len} & \textbf{Residues} & \textbf{Qubits} & \textbf{Min E} & \textbf{Max E} & \textbf{Range} & \textbf{Time (s)} & \textbf{Depth} \\
\hline
4aoi~\cite{cui2012discovery} & VVLPYMKHGDLRNF & 14 & 1155--1168 & 102 & 23245.37 & 32378.95 & 9133.58 & 13328.65 & 413 \\
4cig~\cite{peat2014interrogating} & VRDQAEHLKTAVQM & 14 & 165--178   & 102 & 21375.59 & 29846.54 & 8470.94 & 17293.54 & 413 \\
4tmk~\cite{lavie1998structural} & IEGLEGAGKTTARN & 14 & 8--21      & 102 & 22590.21 & 29135.42 & 6545.21 & 199292.66 & 413 \\
3d7z~\cite{angell2008biphenyl} & YLVTHLMGADLNNI & 14 & 103--116   & 102 & 22979.86 & 29707.30 & 6727.43 & 156289.48 & 413 \\
4jpx~\cite{ronau2013additional} & DYLEAYGKGGVKA  & 13 & 154--166   &  92 & 16962.09 & 22231.95 & 5269.86 & 90422.62  & 373 \\
1yc4~\cite{kreusch2005crystal} & ELISNSSDALDKI  & 13 & 47--59     &  92 & 16129.38 & 20745.81 & 4616.42 & 15777.29  & 373 \\
5cqu~\cite{swider2015synthesis} & RKLGRGKYSEVFE  & 13 & 43--55     &  92 & 17865.39 & 22801.51 & 4936.12 & 7620.94   & 373 \\
2qbs~\cite{wilson2007structure} & HCSAGIGRSGT    & 11 & 214--224   &  72 & 6691.57  & 9356.87  & 2665.30 & 13899.11  & 293 \\
4f5y~\cite{shang2012crystal} & GLAWSYYIGYL    & 11 & 158--168   &  72 & 6408.50  & 8858.60  & 2450.10 & 6157.46   & 293 \\
5cxa~\cite{rouanet2017zinc} & FDGKGGILAHA    & 11 & 174--184   &  72 & 6946.43  & 9298.82  & 2352.40 & 5638.71   & 293 \\
4mc1~\cite{ganguly2014structural} & LLDTGADDTV     & 10 & 23--32     &  63 & 4092.24  & 6199.23  & 2106.99 & 5609.02   & 257 \\
3d83~\cite{angell2008biphenyl} & YLVTHLMGAD     & 10 & 103--112   &  63 & 4235.34  & 6119.16  & 1883.82 & 19833.57  & 257 \\
3vf7~\cite{chang2012potent} & LLDTGADDTV     & 10 & 23--32     &  63 & 3975.02  & 6162.42  & 2187.40 & 5348.25   & 257 \\
4y79~\cite{chan2007factor} & DACQGDSGG      &  9 & 189--197   &  54 & 1549.16  & 2874.21  & 1325.05 & 207445.70 & 221 \\
1e2l~\cite{prota2000kinetics} & AQITMGMPY      &  9 & 124--132   &  54 & 1509.66  & 2837.82  & 1328.15 & 12951.69  & 221 \\
1gx8~\cite{kontopidis2002ligand} & SAPLRVYVE      &  9 & 36--44     &  54 & 1626.02  & 3053.53  & 1427.51 & 14080.77  & 221 \\
1m7y~\cite{capitani2002apple} & TAGATSANE      &  9 & 117--125   &  54 & 1420.38  & 2714.98  & 1294.60 & 12918.04  & 221 \\
3dx3~\cite{kuntz2009molecular} & HNDPGWI        &  7 & 90--96     &  38 & 339.99   & 962.62   & 622.63  & 4661.24   & 157 \\
5c28~\cite{shipe2015discovery} & CDLCSVT        &  7 & 663--669   &  38 & 386.81   & 792.78   & 405.97  & 114029.96 & 157 \\
3ibi~\cite{vitale2009carbonic} & IQFHFH         &  6 & 91--96     &  23 & 120.66   & 455.42   & 334.76  & 4486.62   & 97  \\
2v25~\cite{muller2007bacterial} & ATFTIT         &  6 & 81--86     &  23 & 100.42   & 340.83   & 240.42  & 22356.46  & 97  \\
3ckz~\cite{collins2008crystal} & VKDRS          &  5 & 149--153   &  12 & 10.43    & 14.65    & 4.22    & 5763.36   & 53  \\
3eax~\cite{liu2008targeting} & RYRDV          &  5 & 45--49     &  12 & 10.36    & 16.02    & 5.66    & 4028.72   & 53  \\
\hline
\end{tabular}
\end{table}

\subsection*{Structural Evaluation and Docking Consistency}

To comprehensively evaluate the practical performance of our framework in protein structure prediction tasks, we conducted a systematic analysis of the predicted structures from two perspectives: geometric accuracy and functional consistency. Geometric accuracy was assessed using the RMSD (Root-Mean-Square Deviation), which measures the structural proximity between the quantum-predicted and experimentally determined conformations. Functional consistency was evaluated based on AutoDock scoring functions, comparing the binding affinity of docking simulations using different predicted structures.

Our RMSD-based geometric comparison across 23 protein binding site fragments indicates that the quantum method achieved lower RMSD values in 18 samples, outperforming AlphaFold3 (AF3), which had better results in the remaining 5 cases (Figure~\ref{fig:rmsd_bar}). The average RMSD for quantum predictions was 3.33~\AA, with a median of 3.53~\AA, while AF3 predictions yielded an average RMSD of 3.87~\AA\ and a median of 3.92~\AA. Notably, the quantum method exhibited stronger structural fidelity in shorter sequences (5-9 residues) and in regions with higher structural flexibility. The predicted structures were used as ligands in docking experiments with a standardized receptor structure, and binding affinity were recorded over multiple runs. Results demonstrate that quantum-predicted structures achieved better binding affinities in 21 out of 23 cases (Figure~\ref{fig:affinity_bar}). The average binding affinity for the quantum method was -4.38 kcal/mol, while the AF3-predicted structures averaged -4.00 kcal/mol. This suggests that quantum predictions not only exhibit superior geometric precision, but also more accurately capture the energetic characteristics of real biomolecular interactions. Furthermore, a cross-analysis of structural and functional metrics reveals that quantum-predicted structures exhibit a strong alignment between geometric and energetic consistency. Specifically, conformations with lower RMSD values tend to display more favorable binding affinities. This relationship is quantitatively supported by Pearson ($-0.836$) and Spearman ($-0.823$) coefficients (see Supplementary Figure S10), confirming a statistically significant correlation between RMSD and docking affinity across the tested fragments. This pattern suggests that, within the current problem scale, quantum predictions effectively capture both low-energy conformations and biologically relevant binding configurations. In summary, the quantum computing framework presented in this study delivers structural outputs on real quantum hardware that outperform state-of-the-art deep learning methods in both geometric and functional aspects. Particularly in low-information or small-sample prediction scenarios, where traditional models tend to degrade, our framework demonstrates significant advantages, highlighting its potential utility and scalability in structural biology and bioengineering applications.

In addition to the comparison with AlphaFold3, we further benchmarked our quantum framework against classical simulation--based structure prediction methods to examine its geometric and energetic advantages more broadly. As shown in Supplementary Figures~S3 and S4, the quantum predictions consistently produced lower RMSD and more favorable docking affinities than their classical counterparts. Across the 23 tested fragments, the quantum RMSD averaged 3.68~\AA\ (median 3.54~\AA) versus 6.34~\AA\ (median 6.12~\AA) for classical models, reflecting a 42\% improvement in geometric accuracy. Correspondingly, the mean binding affinity improved from --3.48~kcal~mol$^{-1}$ (classical) to --4.52~kcal~mol$^{-1}$ (quantum), with 21 of 23 samples exhibiting lower (better) docking free energies. The most significant geometric gains were observed in flexible or short-chain fragments such as \textit{2qbs}, \textit{3ckz}, and \textit{4tmk}, where quantum-predicted backbones maintained native-like conformations without excessive over-compaction typical of classical energy-minimization approaches. Energetically, cases like \textit{4aoi}, \textit{4tmk}, and \textit{4y79} demonstrated binding-free-energy reductions exceeding 1~kcal~mol$^{-1}$, suggesting that the quantum ansatz more accurately encodes local electronic interactions relevant to ligand recognition. These trends together indicate that the proposed quantum--variational framework not only enhances structural fidelity but also captures the physical realism of binding energetics, offering a coherent description of biomolecular conformations that classical predictors often approximate empirically.

\subsection*{Case Studies in Real Applications}
To evaluate the practical usability of our quantum computing framework in real-world tasks, we conducted a set of preliminary experiments. Although solving a complete drug discovery pipeline with the current framework remains challenging, it is already feasible to test selected docking-relevant fragments from pharmaceutically meaningful examples to verify its practical performance.
We selected a total of seven real protein fragments with potential applications in drug discovery as validation examples for practical use cases, in order to assess the potential of the current framework in real-world drug discovery problems. All fragments were derived from different protein structures. The sequence "YAGYS" originates from the protein with ID 6mu3, located in the A chain, residues 91-95. This protein is Anti-HIV-1 Fab 2G12, a broadly neutralizing monoclonal antibody targeting the HIV-1 viral envelope glycoprotein. It is widely used in anti-HIV vaccine development and therapeutic antibody research~\cite{chen2009human}. The sequence "DWGGM" comes from the protein with ID 3ans, located in the A chain, residues 335-339. This protein is Human soluble epoxide hydrolase (sEH), an enzyme involved in lipid metabolism. It is closely associated with diseases such as inflammation, cardiovascular disorders, and chronic pain, making it a key target for developing anti-inflammatory and analgesic drugs~\cite{morisseau2013impact}. The sequence "IHGIGGFI" is derived from the protein with ID 1a9m, located in chain A, residues 47-54. This protein is HIV-1 protease G48H, a mutant of HIV-1 protease associated with antiretroviral drug resistance. Studying its structure and function is critical for developing new HIV therapies~\cite{hong1997structure}. The sequence "DGKMKGLAF" originates from the protein with ID 1qin, located in chain B, residues 154-162. This protein is lactoylglutathione lyase, also known as glyoxalase I, which plays a potential role in oxidative stress and glycation-related diseases, such as complications from diabetes. It serves as a potential target for developing drugs to treat diabetic complications~\cite{korithoski2007involvement}. The sequence "NNLGTIAKSGT" comes from the protein with ID 3b26, located in chain A, residues 98-108. This protein is Heat Shock Protein HSP 90-alpha, a molecular chaperone involved in protein folding, stabilization, and degradation. It is extensively implicated in pathological processes, including cancer and neurodegenerative diseases, and is an important target for anti-cancer drug development~\cite{shi2014plasma}. The sequence "KSIVDSGTTNLR" is derived from the protein with ID 1fkn, located in chain A, residues 224-235. This protein is Beta-Secretase (BACE1), a critical enzyme in Alzheimer’s disease research. The structure of its complex with inhibitors provides significant insights for developing anti-Alzheimer’s drugs that reduce $\beta$-amyloid production~\cite{yan2014targeting}. The sequence "GAVEDGATMTFF" originates from the protein with ID 2xxx, located in chain A, residues 136-147. This protein is GLUATAMATE RECEPTOR, IONOTROPIC KAINATE 2 (GluK2), an ionotropic glutamate receptor critical for neuronal signal transmission and synaptic plasticity. It is closely associated with epilepsy, neurodegenerative diseases, and mental disorders, making it an essential target for developing treatments for neurological diseases~\cite{peret2014contribution}. The selected fragments have potential as drug targets, key sites for inhibitors, or for revealing pathological mechanisms in the fields of drug discovery and disease treatment. The quantum method's predicted structures were refined using standard amino acid templates to add atomic positions, where the predicted vectors represented directions between adjacent C$\alpha$ atoms~\cite{samudrala1998all}. For comparison, we used AlphaFold3 to predict the same sequences and processed the results following the same pipeline for charge and hydrogen atom addition to ensure consistency in docking tests (Figure~\ref{fig1}B). We conducted docking tests using AutoDock Vina to evaluate the structures predicted by the quantum method and AlphaFold3~\cite{eberhardt2021autodock,abramson2024accurate}. AutoDock Vina employs classical search algorithms to perform multiple docking attempts for each protein-ligand complex, ultimately generating multiple optimal results. Each result was evaluated using three metrics: affinity, and the lower and upper bounds of the root-mean-square deviation (RMSD). Affinity provides an intuitive measure of docking performance, while the RMSD bounds assess structural variability~\cite{lopez2016new}. To ensure the reliability of the results, we conducted multiple docking experiments for structures predicted by both methods. Each docking test used the same random seed to eliminate stochastic effects and ensure reproducibility. The average values across multiple experiments were used to compare the overall performance of the quantum method and AlphaFold3.

\begin{figure*}
\centering
\includegraphics[width=1\linewidth]{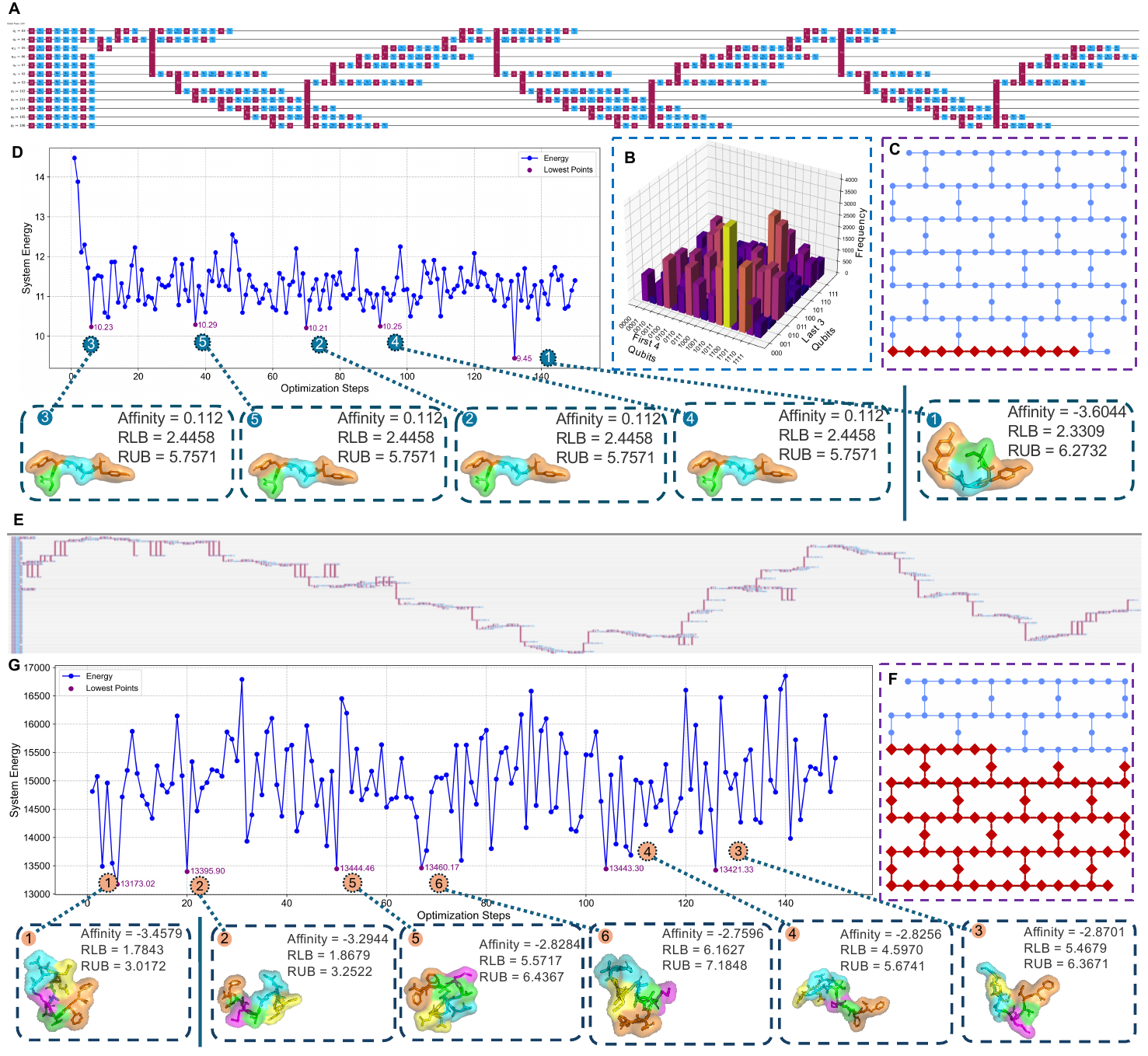}
\caption{\textbf{Details of two example cases in real-world application scenarios.}
(\textbf{A–D}) Case1, 6mu3 (YAGYS): (A) The quantum circuit composed of native gates of the quantum processor, generated after compiling the modeled problem for Case 1.  
(B) The distribution plot of the measured quantum states, where the tallest bar represents the most frequently observed outcome. This highest-probability result corresponds to the final expected measurement outcome. 
(C) The actual qubits occupied by Case 1 on the quantum processor.  
(D) VQE energy trace with five lowest-energy conformers (insets: docking affinity, RLB, RUB). 
(\textbf{E–G})Case2, 2xxx (GAVEDGATMTFF): (E) The quantum circuit composed of native gates of the quantum processor, generated after compiling the modeled problem for Case 2.  
(F) The actual qubits occupied by Case 2 on the quantum processor.  
(G) VQE energy trace with six lowest-energy conformers (insets: docking metrics).}
\label{fig2}
\end{figure*}

Among the seven cases described above, we selected two representative examples: 6mu3 and 2xxx to visualize their quantum-system energy landscapes and to illustrate, within the practical engineering framework, the corresponding resource usage and native-gate quantum circuits. The two examples were chosen for the following reasons. First, they lie at opposite ends of the sequence-length spectrum—one peptide contains five residues and the other twelve—so their qubit requirements, circuit depths, and total energy scales differ markedly, providing a clear basis for analysing how system characteristics change with problem size. Second, the patterns of inter-residue interactions in the two sequences are entirely different, yielding substantially different Hamiltonians; this contrast highlights the versatility and general applicability of our framework. In the case of 6mu3, shown in Figure~\ref{fig2}A-D, we selected the five lowest-energy conformations for visualization. The sequence numbers here represent the ranking starting from the lowest energy. For horizontal comparison, we conducted docking tests using the conformations corresponding to a few other low-energy states, excluding the ground-state energy conformation, with the same receptor molecule in the dataset. During testing, we performed 20 docking tests for each predicted structure, using a different random seed for each test. Each test retained 9 result sets, yielding a total of 180 possible docking outcomes. The final docking score was calculated by averaging three metrics—affinity, RLB (RMSD lower bound), and RUB (RMSD upper bound)—across all 180 results, forming a three-dimensional scoring system. In this case, the energy differences for conformations ranked 2–4 were within the range of ±0.05, and their structures were identical after reverse mapping. These energy differences were caused by perturbations in the quantum system. However, the error magnitude was within a percentage level, so it did not affect solving practical problems. The affinity of these four conformations with the same receptor molecule was greater than 0, indicating that they could not dock with the receptor molecule. The conformation corresponding to the lowest energy point successfully docked, validating the rationality of the method. In the case of 2xxx, shown in Figure~\ref{fig2}E-G, we selected six conformations corresponding to the lowest system energies. Owing to the increased number of amino acids in the second instance, the energy span of the quantum system increased significantly, which resulted in these six conformations being distinct from one another. We applied the same testing method used in the example shown in Figure~\ref{fig2}A-D for horizontal comparison on this set of results. In the tests, a clear positive correlation was observed between the quantum system energy of the conformations and their docking affinity. Specifically, as the quantum system energy of the conformations decreased, the resulting conformations became increasingly suitable for docking and closer to the actual docking structure.

\begin{figure*}
    \centering
    \includegraphics[width=0.9\linewidth]{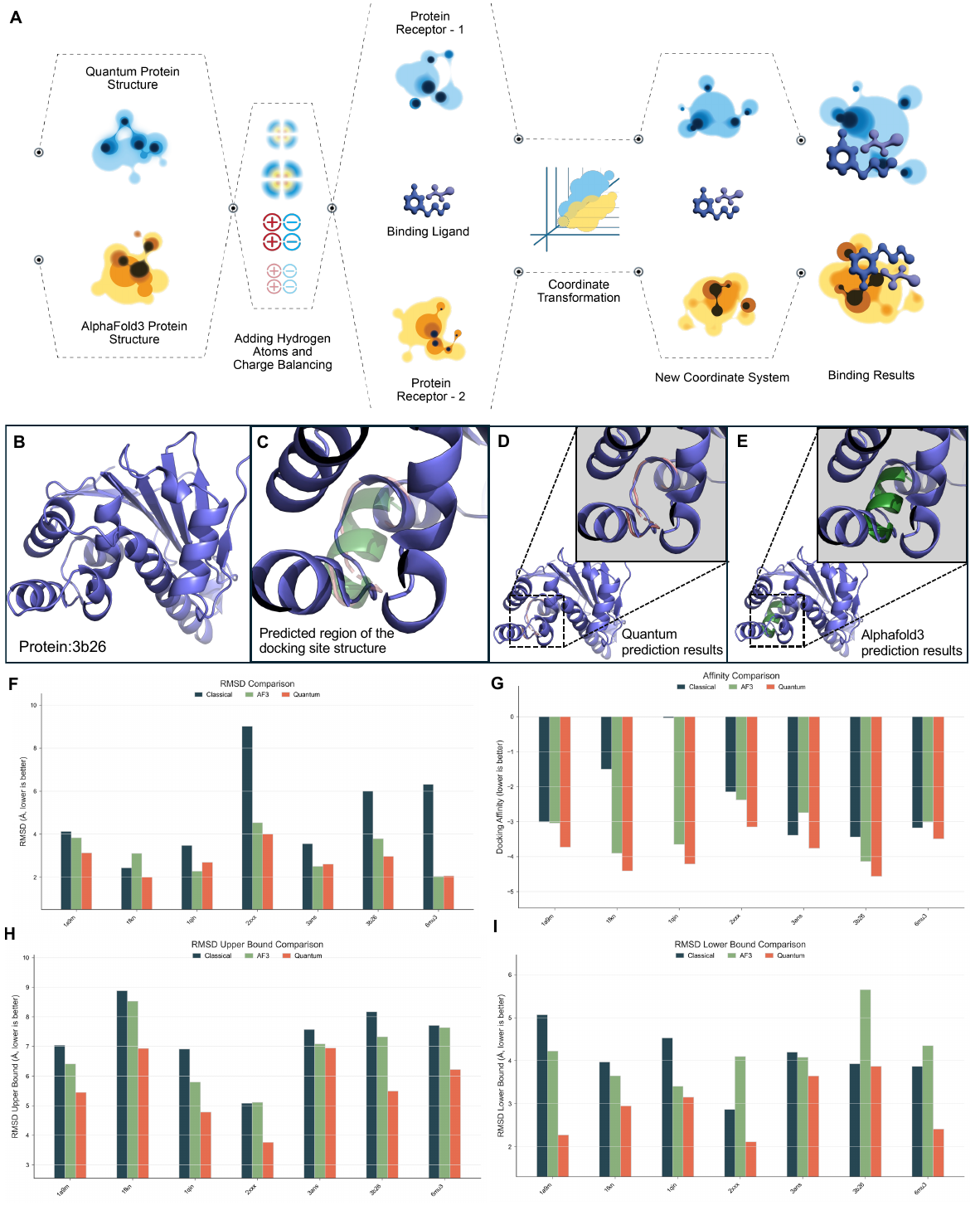}
    \caption{\textbf{Structural testing and visual representation}
    (\textbf{A}) Docking workflow. Predicted structures from both methods are hydrogenated, charge-neutralized, and docked against the same receptor using an identical postprocessing pipeline. 
(\textbf{B}) Native structure of 3b26. 
(\textbf{C}) Docking pocket view. 
(\textbf{D}) Quantum overlay. The quantum prediction accurately follows the fragment trajectory and backbone conformation, showing high consistency with the reference structure. 
(\textbf{E}) AF3 overlay. The AF3 prediction tends to converge to a local minimum under limited sequence information. 
(\textbf{F}) RMSD comparison. Structural deviation between predicted and reference structures; smaller RMSD indicates higher accuracy. 
(\textbf{G}) Affinity comparison. 
(\textbf{H}) RMSD upper bound. 
(\textbf{I}) RMSD lower bound.
}
    \label{fig3}
\end{figure*}

Compared to AlphaFold3 predictions, the quantum method's predicted structures achieved approximately 20\% reductions in the upper and lower bounds of RMSD in different test cases during docking tests, indicating higher consistency and accuracy in structural prediction (Figure~\ref{fig3}H-I). Furthermore, compared to the classical minimization baseline, the quantum method achieved an even larger improvement, with average RMSD upper- and lower-bound reductions of 35–45\% across all cases, demonstrating that quantum optimization can escape shallow local minima that often trap classical gradient-based refinement. In terms of affinity score, the quantum method achieved an average improvement of about 10\% over AlphaFold3 (Figure~\ref{fig3}G). Relative to the classical pipeline, the quantum predictions yielded consistently stronger docking affinities (on average by 0.8–1.0~kcal/mol), indicating more favorable energetic stability of the predicted binding conformations. Additionally, the quantum method exhibited a tighter distribution of affinity scores, with significantly fewer outliers, further validating its reliability in optimizing docking performance. 

Across all benchmark fragments, the quantum framework achieved the lowest RMSD values in five out of seven test cases when compared simultaneously against both AlphaFold3 and classical methods (Figure~\ref{fig3}F). For example, in 1a9m, 3b26, and 3ans, quantum predictions achieved RMSD reductions of 25–40\% relative to the classical minimization and 15–20\% relative to AF3, reflecting a closer recovery of experimental structures. Only in 6mu3, due to the short peptide length and limited conformational flexibility, did all methods converge to similar results within a 0.1~Å difference.

For protein structure prediction, besides evaluating predictive accuracy in docking tasks through binding affinity scores and RMSD bounds of docking poses, another important metric is the RMSD itself. The RMSD (root-mean-square deviation) metric is used to quantify structural differences between the predicted and experimental structures. In our experiments, we compared the RMSD of structures predicted by the VQE method and those predicted by AF3 against the corresponding experimental structures in the dataset. As shown in Figure~\ref{fig3}F, the VQE method outperformed AF3 in six out of seven cases. When the classical baseline is also considered, the overall ranking of RMSD performance follows: Quantum $<$ AF3 $<$ Classical, confirming the progressive improvement from classical to quantum methods. For 6mu3, due to its short chain length and the limited degrees of freedom of the modeling approach, it lagged behind AF3 by only a percentage-level margin. Although most experimental results did not significantly surpass AF3 predictions due to limitations in modeling precision, they nevertheless demonstrate that quantum methods possess certain advantages.

Here we select a representative experimental example to further illustrate the results. As shown in Figure~\ref{fig3}B-E, we chose a fragment from the binding pocket region of protein 3b26 for demonstration. In our experiments, this fragment consists of 11 amino acids—one of the longer test cases—which facilitates clear visualization. In the visualized outcome, the structure predicted by the quantum method clearly and accurately captures the overall fold and key structural features of the selected segment, showing a high degree of similarity to the experimentally determined structure in the dataset. In contrast, the AF3 prediction shown in Figure E deviates substantially from the experimental result: AF3 models this region as an $\alpha$-helical secondary structure, which differs markedly from the true conformation and loses essential structural characteristics. The classical model, by contrast, produces a highly distorted conformation that fails to preserve backbone geometry or side-chain orientation, consistent with its higher RMSD and weaker docking affinity (Figure~\ref{fig3}F–G). In our experiments, many AF3 predictions tended to default to helical structures for unknown regions. We hypothesize that this bias arises because helices have high recognizability and pronounced patterns, which create an information “pit” in pattern-recognition models; under limited system information capacity, the deep learning model gravitates toward predictions with the most distinctive features, thus sacrificing accuracy. In contrast, quantum methods build from first principles and do not fall into pattern-recognition traps. Even when system information capacity is severely constrained, they can still produce results with reliable fidelity.

\subsection*{Scalability}
\begin{figure*}
    \centering
    \includegraphics[width=\linewidth]{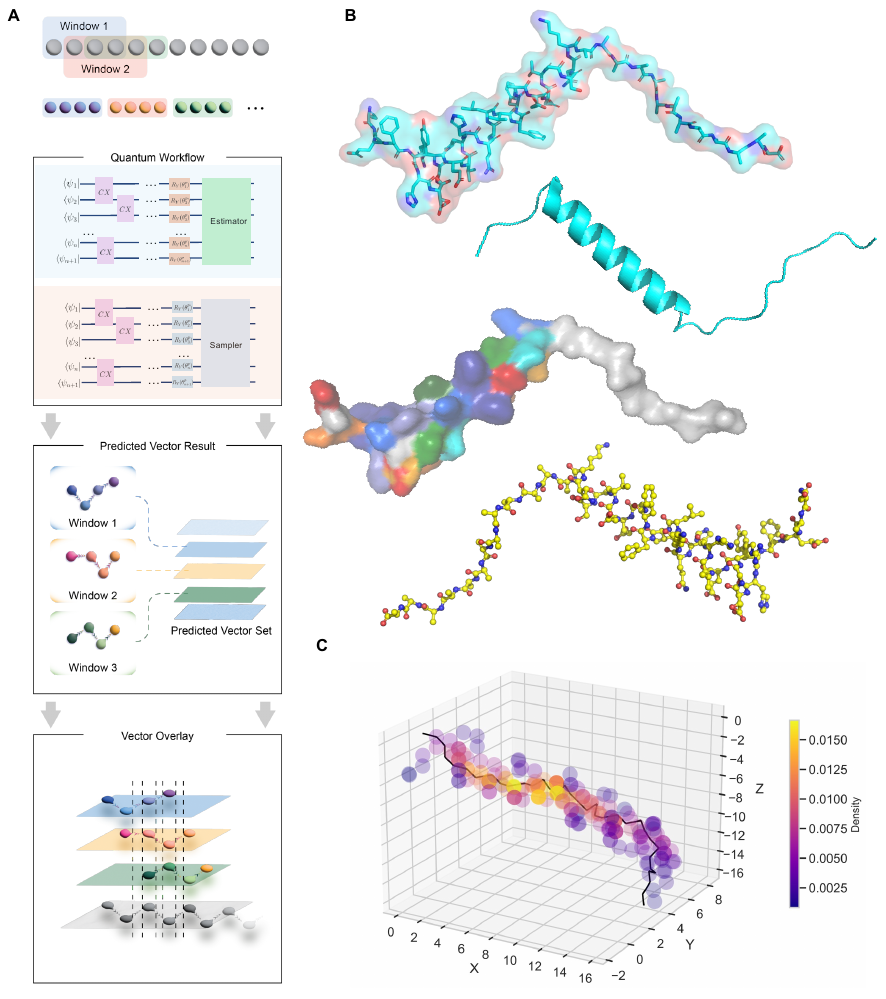}
    \caption{\textbf{Scalability test of the prediction framework. Sliding-window quantum prediction of A$\beta$42. }
    (\textbf{A}) Segmentation and two-stage VQE workflow (7-residue window, one-residue stride) Repeatedly invoking the framework to predict multiple fragments and then integrating them into a complete protein structure.  
    (\textbf{B}) Reassembled full-length A$\beta$42 model shown in surface, cartoon, and ball-and-stick views.  
    (\textbf{C}) Backbone vector density map; high-density regions mark the most probable conformations. Lighter-colored regions indicate lower variance across multiple predictions, suggesting that repeated invocations did not introduce significant deviations in the final structure. The overall prediction error remains within an acceptable range.}
    \label{fig-Abeta}
\end{figure*}

Due to the current limitations of quantum resources, the quantum method can predict protein structures composed of sequences of 10 to 15 amino acids at most~\cite{bhuvaneswari2023computational}. Longer sequences would exceed quantum resource limits and face increased noise interference. To address this issue, we developed a sliding-window model to predict longer amino acid sequences~\cite{chen2006optimization}. After experimentally testing different combinations of step sizes and window sizes, we selected a configuration with a step size of 1 and a window size of 7 (Figure~\ref{fig-Abeta}). In selecting the sliding window size and stride, we primarily considered the following factors: the information capacity of the quantum system, the accuracy of the prediction process, and the rationality of the structural predictions. Choosing a window size of 7 nodes utilizes over 70\% of the qubits in currently available quantum processors and allows for considering the interactions between more adjacent nodes, thereby avoiding a "locally optimal" structure that could reduce the overall structural accuracy. Using a larger window would burden the system's information capacity, and without sufficient additional qubits, the circuit depth would increase significantly, leading to excessive computational costs. Conversely, a smaller window would lower the overall reliability of the structure. A stride of 1 was chosen to maximize information density, improving the reliability of the predicted protein structure by increasing the number of computation rounds. Each windowed amino acid segment was modeled and predicted using an independent workflow, with 30 optimization iterations performed for each segment to identify the ground state energy. Predictions from different segments were combined into a larger vector set representing the extension directions between protein nodes. Each vector was expressed as 7 sub-vectors, which were then combined using weighted averaging to compute the final direction vectors between adjacent nodes, ultimately generating the complete A$\beta$ protein structure~\cite{cheng2013swfoldrate}. To produce a clearer and more accurate final structure, we refined the backbone generated by the quantum method by comparing it with the secondary structure predictions from AF3, yielding the final model.

\section*{Discussion}
This study provides the validation of the feasibility and potential of quantum computing, particularly the Variational Quantum Eigensolver (VQE), in predicting protein structures. By mapping the protein conformation problem into the Hamiltonian optimization problem of a quantum system, quantum computing demonstrates unique advantages in handling molecular energy domains. Through parameterized quantum circuits, quantum systems efficiently generate energy values and iteratively refine them using classical optimizers, rapidly converging to the lowest-energy conformation~\cite{wolynes2015evolution}. In comparison to existing AI methods such as deep learning, quantum methods offer higher theoretical reliability. While AI-based approaches, such as AlphaFold3~\cite{abramson2024accurate}, achieve exceptional efficiency in predicting new sequences after model training, they are inherently probabilistic models reliant on large-scale data and suffer from limited interpretability~\cite{gomes2022protein}. These characteristics may result in limitations when addressing local structures or specific cases. In docking tasks involving small-scale fragments similar to those we used, the focus is on interface pocket surfaces formed by short peptide chains that require highly accurate structural predictions. For deep learning models like AlphaFold3~\cite{abramson2024accurate}, although these models are built with neural networks containing millions of parameters, the information capacity for predicting such structures is extremely limited. When predicting these structures, the models are prone to falling into "information traps," such as generating structures that appear complete and meet basic amino acid distance characteristics but fail to satisfy amino acid residue interactions or chirality constraints. As a result, these structures do not meet the requirements of docking tasks. In contrast, quantum methods model protein conformations from the foundation of physical principles, circumventing the probabilistic constraints of AI methods and providing theoretically rigorous predictions. Additionally, quantum methods avoid the high computational costs associated with classical approaches, redefining protein structure prediction problems from a high-dimensional perspective.

As protein sequences lengthen and the number of interactions increases, the corresponding Hamiltonian grows more complex, and the depth of the quantum circuit also increases. Consequently, noise accumulation becomes more pronounced, which can slow VQE convergence, trap the system in local minima, or shift the estimated energy away from its true value, all of which undermine the accuracy of the protein backbone conformation. Meanwhile, if the number of measurements in the final sampling stage is insufficient, statistical (shot) noise remains large; if measurements are too frequent, hardware is required to maintain the same circuit state for a longer duration, possibly amplifying noise further~\cite{oliv2022evaluating}. Some studies have demonstrated that using coarse-grained or lower-dimensional energy functions to represent protein structures can reduce the reliance on fully atomistic simulations and significantly decrease the sensitivity of quantum systems to noise~\cite{yang2024coupler}. Meanwhile, recent research provides further evidence that non-fault-tolerant quantum devices still possess sufficient performance to address utility tasks~\cite{kim2023evidence,lu2023recent}. In our work, we employed a simplified vector-based protein backbone model, which has been shown to exhibit high robustness~\cite{doga2024perspective,zhang2025qdockbank}, reducing the dimensionality of observables and the overall state space. Although this coarse-grained model cannot perfectly capture atomic-scale interactions, it is sufficient to represent key geometric features and energy trends. Our five model workflow also significantly reduces the impact of noise. In the first stage (estimation), the system avoids large-scale quantum-state measurements and focuses on rapidly estimating total energy, thereby reducing the accumulation of quantum gate and readout errors. In the second stage (sampling), after the circuit parameters have approximately converged, we perform extensive measurements in a single run to fully extract the final quantum-state distribution, applying readout error mitigation techniques to correct for systematic biases in the measurement process. While utility-level quantum devices still exhibit disturbances caused by noise, these disturbances do not have a practical impact on the results or significantly affect the overall performance of the system in the problem scenarios we addressed (Figures~\ref{fig2} and~\ref{fig-Abeta}).

To mitigate noise in quantum processors, previous studies have proposed a range of techniques, such as adding error detection or error correction codes (ECC) to quantum circuits~\cite{krinner2022realizing}, or deploying denoisers at the pulse or measurement level~\cite{fu2024hqc}. While these measures effectively reduce error rates, they also require additional physical qubits and quantum gate operations, increasing hardware demands and runtime~\cite{tepaske2023compressed}. Additionally, zero-noise extrapolation (ZNE) artificially amplifies circuit noise by inserting pairs of quantum gate operations, measures energy values under different noise levels, and extrapolates results toward a "zero-noise" limit ~\cite{giurgica2020digital}. Pulse-level optimizations and dynamic error correction may further suppress decoherence~\cite{he2020zero}. As hardware evolves and quantum error-correcting codes (QEC) mature, gate fidelities and the number of available qubits are expected to increase, providing broader possibilities for larger-scale and higher-accuracy protein structure prediction~\cite{zhang2024high,evered2023high,graham2022multi}. Looking ahead, advancements in quantum hardware, including increased qubit numbers and improved error-correction capabilities, will enable quantum computing to address more complex biological tasks. Potential application areas include long-chain protein structure prediction, modeling multi-molecular interactions, and drug molecule design. Moreover, hybrid approaches combining classical optimizers with deep learning algorithms may further enhance the efficiency and applicability of quantum computing, driving revolutionary advancements in computational biology.

\section*{Methods}

\subsection*{Data Source and Preparation}
The data used in this study were obtained from the PDBbind database~\cite{wang2004pdbbind,wang2005pdbbind,liu2015pdb}, which contains high-quality protein-ligand complex structures and protonation charge data. Protein structures with well-defined active sites were selected to ensure consistent structural information and charge states, thereby standardizing the inputs for the prediction models. Because our approach aims to provide a general solution for protein-ligand docking problems, rather than focusing on any specific drug discovery target, we randomly selected two sets of complexes from the PDBbind database for our experiments. There was no particular rationale for choosing these two sets beyond their availability and completeness; they serve as representative examples to demonstrate the universality of our method. The PDBbind database is widely accepted as a standard resource for evaluating docking and binding-affinity prediction methods and has been used in numerous previous studies~\cite{sajjan2022quantum,rodrigues2025csm,suriana2023enhancing}.

\subsection*{Geometric Encoding on a Tetrahedral Lattice}

A central principle of our encoding strategy is the control of structural granularity when projecting protein conformations into a quantum representation. Earlier lattice-based approaches~\cite{pierri2008lattices,fingerhuth2018quantum,zhou2024variational} typically employed link- or vector-based encodings that preserve both backbone and side-chain variability~\cite{zha2023encoding}. While such fine-grained models capture detailed features, they also lead to steep, rugged energy landscapes that amplify hardware noise and increase the qubit requirements of the quantum system. In this work, we introduce a coarser-grained, single-chain vector representation that omits side-chain degrees of freedom and focuses solely on the peptide backbone. Each amino acid node extends along one of four vectors corresponding to tetrahedral geometry, with unit bond length and a bond angle of approximately $109.4^\circ$, consistent with fundamental stereochemical constraints~\cite{robert2021resource}. By removing side-chain variability, the conformational energy surface mapped into the Hamiltonian becomes smoother, reducing high-frequency fluctuations and yielding a formulation that is more robust to the noise characteristics of current quantum devices. 
The connection vectors between amino acid nodes, inter-residue interactions, and structural constraints are mapped to quantum operators, which together define the system Hamiltonian. The dimensionality of these operators determines the number of qubits required for execution: $n$ qubits span a Hilbert space of dimension $2^n$, and thus problem complexity increases with sequence length and amino acid diversity. Table~\ref{tab:execution} summarizes the number of qubits and circuit depths used in our experiments. For illustration, encoding a residue with four possible extension vectors can be achieved using four qubits, where measurement outcomes such as 0001, 0010, 0100, or 1000 activate one of the four tetrahedral directions~\cite{wang2024efficient}. At initialization, all feasible connection paths for a sequence are encoded in superposition, each equally probable. Variational optimization under the Hamiltonian gradually shifts the probability distribution toward low-energy conformations, where invalid configurations are suppressed by energy penalty terms. Upon measurement, the system collapses to a binary string representing the most probable backbone arrangement, corresponding to the conformation that minimizes the protein’s total energy. Through this granularity-controlled encoding, the Hamiltonian retains the essential geometry of peptide chains while discarding details that would otherwise overwhelm current hardware. This design produces a more tractable optimization landscape and enables stable execution of docking-relevant fragments on real quantum processors.

\subsection*{Hamiltonian}
\paragraph{Turn encoding and indicator projectors.}
For a peptide backbone of length $L$, we allocate $2(L-1)$ qubits, two per turn, to encode the discrete tetrahedral directions. 
Let the two qubits for turn $t\in\{0,\dots,L-2\}$ be indexed as $(t,0)$ and $(t,1)$.
Define single-qubit projectors
\[
P^{(1)}_{t,b}=\frac{I+Z_{t,b}}{2},\qquad 
P^{(0)}_{t,b}=\frac{I-Z_{t,b}}{2}\quad (b\in\{0,1\}),
\]
so that a two-bit direction $d\in\{0,1,2,3\}$ with binary digits $d=(d_0,d_1)$ is selected by the
two-qubit indicator projector
\[
\Pi^{(d)}_{t}\;=\;P^{(d_0)}_{t,0}\,P^{(d_1)}_{t,1}
\;=\;\frac{1}{4}\Big(I+s^{(d)}_{t,0}Z_{t,0}\Big)\Big(I+s^{(d)}_{t,1}Z_{t,1}\Big),
\]
where $s^{(d)}_{t,b}=+1$ if $d_b=1$ and $s^{(d)}_{t,b}=-1$ if $d_b=0$.
By construction,
$\sum_{d=0}^{3}\Pi^{(d)}_{t}=I$ and $\Pi^{(d)}_{t}\Pi^{(e)}_{t}=\delta_{de}\Pi^{(d)}_{t}$.
In code, these operators are implemented with \texttt{SparsePauliOp} by composing
$\frac{1}{2}(I\pm Z)$ factors for the appropriate qubit indices (see \texttt{\_build\_vector\_qubit} and
\texttt{\_build\_turn\_indicator\_fun}).

\paragraph{Total Hamiltonian.}
The overall Hamiltonian combines all stereochemical, geometric, steric, and interaction constraints:
\[
H_{\mathrm{t}}\;=\;\lambda_c H_{\mathrm{c}}+\lambda_g H_{\mathrm{g}}+\lambda_d H_{\mathrm{d}}+\lambda_i H_{\mathrm{i}},
\qquad \lambda_\bullet>0.
\]
Each term reduces to a sparse real linear combination of Pauli strings
\[
H_{\mathrm{t}}=\sum_{\ell}\gamma_\ell\,P_\ell,\qquad P_\ell\in\{I,Z\}^{\otimes n},\;\gamma_\ell\in\mathbb{R},
\]
i.e., an Ising-type Hamiltonian with $Z$-only support, stored compactly in \texttt{SparsePauliOp}.
This guarantees that all expectation values can be estimated directly in the computational basis.

\paragraph{(1) Geometric one-hot and turn-consistency constraints.}
We enforce that each turn takes exactly one of the four admissible directions and that successive
turns respect lattice geometry (bond length and $\sim109.4^\circ$ angle) via
\[
H_{\mathrm{g}} \;=\; \mu_{\mathrm{oh}}\sum_{t}\Big(1-\sum_{d=0}^{3}\Pi^{(d)}_{t}\Big)^{\!2}
\;+\;\mu_{\mathrm{ang}}\sum_{t=1}^{L-2}\sum_{(d,e)\notin\mathcal{A}}
\Pi^{(d)}_{t-1}\,\Pi^{(e)}_{t},
\]
where $\mathcal{A}$ is the set of allowed direction pairs for adjacent turns
under tetrahedral geometry. 
Expanding the one-hot penalty gives only $I, Z_{t,0}, Z_{t,1}, Z_{t,0}Z_{t,1}$ terms.
The pairwise angle penalties yield at most 4-body $Z$ products.

\paragraph{(2) Chirality and backtracking penalties.}
To ensure stereochemical correctness and avoid immediate reversals, we penalize forbidden
turn pairs and triples:
\[
H_{\mathrm{c}}\;=\;\kappa_{\mathrm{bt}}\sum_{t=1}^{L-2}\sum_{(d,e)\in\mathcal{F}_{\mathrm{bt}}}
\Pi^{(d)}_{t-1}\,\Pi^{(e)}_{t}
\;+\;
\kappa_{\chi}\sum_{t=1}^{L-3}\sum_{(d,e,f)\in\mathcal{F}_{\chi}}
\Pi^{(d)}_{t-1}\,\Pi^{(e)}_{t}\,\Pi^{(f)}_{t+1}.
\]
Each triple product expands into up to 6-body $Z$ strings, but remains sparse.

\paragraph{(3) Distance/steric exclusion.}
Self-avoidance and steric clashes are encoded via a collision mask $C^{(d,e)}_{t,u}$:
\[
H_{\mathrm{d}}\;=\;\eta\sum_{0\le t<u\le L-2}\;\sum_{d,e=0}^{3}
C^{(d,e)}_{t,u}\;\Pi^{(d)}_{t}\,\Pi^{(e)}_{u}.
\]
This produces sums of 4-body $Z$ strings that penalize invalid spatial contacts.
\paragraph{(4) Pairwise interaction energies.}
Residue-residue stabilization is incorporated by a contact mask $M^{(d,e)}_{t,u}$ and
interaction matrix $J_{ab}$:
\[
H_{\mathrm{i}}\;=\;\sum_{0\le t<u\le L-2}\;\sum_{d,e=0}^{3}
J_{a_t a_u}\;M^{(d,e)}_{t,u}\;\Pi^{(d)}_{t}\,\Pi^{(e)}_{u}.
\]

\paragraph{Representative Pauli-$Z$ expansion.}
Writing $\Pi^{(d)}_{t}=\frac{1}{4}\sum_{S\subseteq\{(t,0),(t,1)\}} 
\alpha^{(d)}_{S}\,Z_{S}$ with $Z_{\emptyset}=I$, $Z_{\{(t,b)\}}=Z_{t,b}$ and
$Z_{\{(t,0),(t,1)\}}=Z_{t,0}Z_{t,1}$, any quadratic form $\Pi^{(d)}_{t}\Pi^{(e)}_{u}$ expands as
\[
\Pi^{(d)}_{t}\Pi^{(e)}_{u}=\frac{1}{16}\!\!\sum_{S,T}
\alpha^{(d)}_{S}\,\alpha^{(e)}_{T}\; Z_{S}\,Z_{T},
\]
containing only $I$ and $Z$ products. 
Thus the full Hamiltonian is an Ising Hamiltonian with sparse Pauli-$Z$ support,
directly implementable on utility-level hardware.
\paragraph{Ansatz and optimization.}
We solve $\min_{\theta}\,\langle\psi(\theta)|H_{\mathrm{t}}|\psi(\theta)\rangle$ using VQE,
with an \texttt{EfficientSU2} ansatz and COBYLA optimizer.
Circuit depths and qubit counts for all fragments are summarized in Table~\ref{tab:execution}.

\subsection*{Quantum Circuit Design and Variational Optimization}
In realistic scenarios, the qubit requirement grows significantly with the amino acid sequence length. For very short sequences, fewer than 10 qubits may suffice. However, in our experiments, when dealing with 12 amino acids, at least 77 qubits are needed to encode all relevant operators. Moreover, to satisfy certain coupling conditions and allow for additional SWAP operations that reduce overall circuit depth, we reserve about 5 extra qubits (totaling 82) as a buffer. This strategy effectively lowers the entangling-gate overhead (see, e.g.,~\cite{li2020qubit,li2019tackling} for related approaches). 
In many recent studies, the \texttt{EfficientSU2} ansatz has emerged as a widely used and robust parameterized circuit, recognized for its relatively broad applicability and moderate resource requirements~\cite{gultom2024optimizing,dominic2024variational}. To parameterize our circuits, we employ the \textit{EfficientSU2} ansatz from Qiskit. This ansatz typically consists of repeated blocks of single-qubit rotation gates \(R_y\) and \(R_z\), interspersed with CX gates arranged in a circular or linear topology to introduce entanglement. By adjusting the number of repetition layers (\textit{reps}), we can flexibly control the circuit’s expressive power. In a single repetition layer, each qubit undergoes one \(R_yR_z\) rotation, followed by an entangling step that couples adjacent qubits. The theoretical circuit depth of Qiskit's \texttt{EfficientSU2} parameterized quantum circuit grows linearly with the number of qubits \( N \). For our case with \( N = 82 \) qubits: Total theoretical depth is \( 333 \). The actual compiled depth may be different through gate optimization and hardware topology adaptation.
An important advantage of the quantum method is its extremely fast computational speed. Due to the superposition state of the quantum system, during each optimization, the system does not focus on a specific structure for validation like traditional structure-based prediction methods. Instead, the superposition state allows all possible conformations of the current protein structure to exist simultaneously at the physical level, collapsing into a definite ground state as the optimal result through parameter optimization. In our tests, for a sequence with a length of 5 amino acids, the total recorded execution time after 150 iterations was 4079.55 seconds. According to logs from IBM's superconducting quantum computer, the average time per optimization iteration and state measurement, from task submission to result acquisition, was approximately 10 seconds, with about 73.53\% of the time spent on the quantum end. In our setup, the number of shots for optimization tasks was 200, and the number of shots for measurement was 100,000, indicating that the execution efficiency of the quantum computing method is extremely high. The quantum processor itself completes a single task almost instantaneously, with most of the time spent on state preparation and the machine's self-calibration. Additionally, part of the time was limited by the performance of the classical computing backend. Under computational models of equivalent accuracy, the quantum method is several orders of magnitude faster than all current methods based on classical computing.

\subsection*{Docking Evaluation with AutoDock Vina}

To evaluate the structural quality and potential functional effectiveness of the predicted protein fragment conformations, we conducted docking-based validation using AutoDock Vina. To ensure fairness and reproducibility, all structures—those predicted by our quantum computing framework and those generated by AlphaFold3 (AF3)—were subjected to a standardized preprocessing and docking pipeline as follows: 

\textbf{Structure preprocessing.} All predicted structures were preprocessed using AutoDockTools before docking: polar hydrogens were added, Gasteiger charges were assigned, and backbone conformations were preserved. Receptor structures were derived from native PDBbind entries, and the ligands were held constant across all tests. The goal was to evaluate how different predicted conformations influence binding capability with the same ligand.

\textbf{Docking protocol.} For each test case, the ligand was docked to the predicted protein fragment using a grid box centered on the native binding site. Each structure underwent 20 independent docking runs, with each run producing 9 docking poses ranked by predicted binding energy, resulting in a total of 180 docking results per structure for statistical analysis.

\textbf{Evaluation metrics.} AutoDock Vina outputs were assessed using three key metrics: (1) Binding affinity score (kcal/mol), where lower scores indicate stronger binding potential; (2) 
RMSD lower bound (RLB), the minimum structural deviation between the top-scoring pose and all others, reflecting prediction stability; and (3) RMSD upper bound (RUB), the maximum deviation among all poses, indicating overall reliability\cite{gallicchio2010binding,lopez2016new}.

\textbf{Summary of results.} All docking results were analyzed statistically across the 20 repeated experiments. As shown in Figure~\ref{fig3}B–D, structures predicted using the quantum method consistently outperformed those from AF3 in terms of both binding affinity and RMSD-based stability, demonstrating greater consistency and accuracy in geometric conformation and functional behavior.

\subsection*{Quantum Hardware and Execution Environment}

All quantum computing tasks in this study were executed on real quantum hardware provided by IBM Quantum, specifically the IBM-Cleveland Clinic quantum processor. This device belongs to IBM's Eagle r3 series within the IBM One line and features high-fidelity gate operations and a relatively stable readout mechanism, making it suitable for small- to medium-scale variational quantum algorithm experiments.

\textbf{Task submission and management.} All tasks were submitted via IBM's Runtime service, using Qiskit’s dynamic scheduling capabilities to automatically allocate backend resources\cite{chow2021ibm}. Each structure prediction task comprised two stages: Stage 1 executes the Variational Quantum Eigensolver (VQE) algorithm to minimize the ground-state energy; Stage 2 recompiles the circuit with fixed optimal parameters and performs large-scale measurements to extract the distribution of the final quantum state. The entire task flow was encapsulated into a two-stage session job and submitted to IBM's job queue for execution.

\textbf{Circuit design and compilation optimization.} To align with the hardware topology of IBM-Cleveland Clinic, we applied gate remapping and gate fusion strategies in our quantum circuit design to minimize circuit depth and the number of two-qubit gates. Additionally, we utilized Qiskit’s transpiler stack to maximize execution fidelity while preserving circuit logic. For key tasks, noise mitigation techniques such as dynamic decoupling and readout error mitigation were employed to reduce systematic errors.

\textbf{Execution time and resource consumption.} Table~\ref{tab:execution} summarizes key statistics for each task, including sequence length, number of qubits used, circuit depth, optimization time, and energy range. Most tasks completed within acceptable time frames, with the most complex jobs requiring no more than 24 hours. It is worth noting that execution time can vary due to IBM’s backend scheduling algorithm: jobs that consume more resources may be placed in lower-priority queues and experience delays while awaiting higher-priority executions. These outliers reflect normal scheduling overheads. As sequence length increases, resource demands (qubit count and circuit depth) grow significantly; however, all tasks remained operationally stable on current utility-level devices\cite{zeng2021simulating}.

\textbf{Hardware limitations and mitigation strategies.} Although today's physical quantum devices are still limited by gate errors, readout noise, and finite coherence times, our two-stage scheduling mechanism and circuit compilation strategies successfully mitigated performance degradation from long-duration runs. This execution framework demonstrates the engineering feasibility of performing structure prediction on present-day utility-level quantum hardware.


\clearpage 

%
\bibliography{reference} 
\bibliographystyle{unsrt}

%
%
%
%
%
%


\section*{Acknowledgments}
\paragraph*{Funding:}
This work was primarily supported by the National Institute on Aging (NIA) under Award Numbers U01AG073323, R21AG083003, R01AG066707, R01AG076448, R01AG082118, R01AG084250, P30AG072959, and RF1AG082211, and the National Institute of Neurological Disorders and Stroke (NINDS) under Award Number RF1NS133812, and Alzheimer's Association (ALZDISCOVERY-1051936) to F.C. This work was also supported in part by NIH grants U01 HG007691, R01 HL155107, and R01 HL166137; by AHA grants AHA957729 and AHA24MERIT1185447; and by EU Horizon 2021 grant 101057619 to J.L. This project has been funded in whole or in part with federal funds from the National Cancer Institute, National Institutes of Health, under contract HHSN261201500003I. The content of this publication does not necessarily reflect the views or policies of the Department of Health and Human Services, nor does mention of trade names, commercial products, or organizations imply endorsement by the U.S. Government. This Research was supported [in part] by the Intramural Research Program of the NIH, National Cancer Institute, Center for Cancer Research. This work was also partially supported by NSF 2238734 and NSF 2311950.
\paragraph*{Author contributions:}
F.C. and Q.G. conceived the study. Y.Z. designed research; Y.Z., Y.Y. and W.M. performed research; J.L., R.N., Y.Z., Y.Y., W.M., Z.W., C.L. and W.J. interpreted the data analysis; Y.Z., Y.Y., W.M., Q.G., F.C., K.L., R.N., and J.L. wrote and critically revised the manuscript. All authors gave final approval of the manuscript.
\paragraph*{Competing interests:}
There are no competing interests to declare.
\paragraph*{Data and materials availability:}
All data, code, and materials used in the analysis are available at Zenodo under the DOI: \url{https://doi.org/10.5281/zenodo.15653560}. This includes the full source code for the quantum structure prediction framework,  Hamiltonian construction scripts, VQE configurations. The repository enables full reproduction of all results presented in this study.


\subsection*{Supplementary materials}
Supplementary Text\\
Tabs. S1\\
Figs. S1 to S10\\


\newpage


\renewcommand{\thefigure}{S\arabic{figure}}
\renewcommand{\thetable}{S\arabic{table}}
\renewcommand{\theequation}{S\arabic{equation}}
\renewcommand{\thepage}{S\arabic{page}}
\setcounter{figure}{0}
\setcounter{table}{0}
\setcounter{equation}{0}
\setcounter{page}{1} 


\begin{center}
\section*{Supplementary Materials for\\ \scititle}

Yuqi Zhang, 
Yuxin Yang, 
William Martin, 
Kingsten Lin, 
Zixu Wang,
Cheng-Chang Lu, 
Weiwen Jiang, 
Ruth Nussinov, 
Joseph Loscalzo,
Qiang Guan$^{\ast}$, 
Feixiong Cheng$^{\ast}$

\small$^\ast$Corresponding author. Email: qguan@kent.edu; chengf@ccf.org.\\

\end{center}

\subsubsection*{This PDF file includes:}
Supplementary Text\\
Table S1\\
Figures S1 to S10\\

\newpage




\subsection*{Coloring Scheme for Protein Visualization in Figure4}

In protein visualization, amino acids are categorized based on their chemical properties, and each category is assigned a specific color to enhance structural understanding. The coloring rules are as follows:
1. \textbf{Hydrophobic Amino Acids}: This group includes alanine (ALA), glycine (GLY), valine (VAL), leucine (LEU), isoleucine (ILE), and proline (PRO). These amino acids tend to be located in the interior of proteins to avoid contact with water. They are colored cyan to represent their hydrophobic nature.
2. \textbf{Negatively Charged Amino Acids}: This group includes aspartic acid (ASP) and glutamic acid (GLU). These amino acids have side chains that carry a negative charge at physiological pH, making them acidic. They are colored yellow to highlight their negative charge.
3. \textbf{Positively Charged Amino Acids}: This group includes arginine (ARG), lysine (LYS), and histidine (HIS). These amino acids have side chains that carry a positive charge at physiological pH, making them basic. They are colored red to represent their positive charge.
4. \textbf{Polar Uncharged Amino Acids}: This group includes serine (SER), threonine (THR), asparagine (ASN), and glutamine (GLN). These amino acids have polar side chains but do not carry a charge at physiological pH. They are colored green to reflect their polarity.
5. \textbf{Sulfur-Containing Amino Acids}: This group includes cysteine (CYS) and methionine (MET). These amino acids contain sulfur atoms and play crucial roles in protein function and structural stability. They are colored magenta to emphasize their unique chemical properties.
6. \textbf{Aromatic Amino Acids}: This group includes phenylalanine (PHE), tyrosine (TYR), and tryptophan (TRP). These amino acids have conjugated aromatic ring structures, which are important for protein stability and interactions. They are colored orange to highlight their aromatic nature.
This coloring scheme is designed to visually distinguish the chemical properties and spatial distribution of amino acids in a protein structure, aiding in the interpretation of protein function and characteristics.

\subsection*{Classical Baseline Pipeline}

To provide a fair reference for performance comparison, we established a classical baseline pipeline that mirrors the structural generation and energy minimization stages of the quantum framework. The entire process consists of three main steps: initial structure generation, structure standardization, and energy minimization. 

First, for each peptide sequence, an initial 3D conformation was constructed using \texttt{PeptideBuilder}, which sequentially connects amino acid residues into an extended backbone model without prior structural bias. Second, the resulting structure was processed through \texttt{PDBFixer} to add missing atoms, reconstruct side chains, and neutralize charges at physiological pH, ensuring chemical completeness and comparability with the quantum-generated outputs. Finally, we performed potential energy minimization using \texttt{OpenMM} with the Amber14 force field and hydrogen-bond constraints under implicit solvent conditions. The minimization employed the \texttt{LangevinMiddleIntegrator} at 300~K and 200 iterations, matching the iteration depth used in the quantum optimization stage. This ensures that both classical and quantum methods operate under equivalent computational budgets for structural refinement.

The classical minimization converges to local energy minima on the predefined potential surface, which is limited by the initial conformational bias and deterministic gradient descent dynamics. In contrast, the quantum approach—based on the Variational Quantum Eigensolver (VQE)—samples from a high-dimensional Hilbert space through parametrized quantum circuits, inherently exploring a larger conformational landscape even within the same iteration count. This allows the quantum framework to avoid premature convergence to shallow minima and to identify configurations that more closely correspond to global or near-global minima in the underlying energy landscape. 

Overall, this classical baseline serves as a physically meaningful yet computationally bounded reference. By maintaining identical iteration limits and postprocessing conditions, the observed advantages of the quantum method—particularly in RMSD reduction and docking affinity improvement—can be attributed to its expanded search capability in the quantum-encoded conformational space rather than to differences in computational effort or sampling time.

\newpage

\begin{table}[htbp]
\centering
\caption{\textbf{Typical hardware parameters of the IBM Eagle (127-qubit) superconducting quantum processor.}}
\label{tab:eagle_specs}
\renewcommand{\arraystretch}{1.25}
\setlength{\tabcolsep}{6pt}
\begin{tabular}{l l l}
\hline
\textbf{Parameter} & \textbf{Typical value} & \textbf{Description} \\
\hline
Processor family & IBM Eagle (r3) & 127-qubit superconducting transmon processor \\
Number of qubits & 127 & Total number of physical qubits \\
Qubit technology & Fixed-frequency transmon & Operated at millikelvin temperatures \\
Topology & Heavy-hex lattice & Nearest-neighbor coupling network \\
Native gates & \{rz, sx, x, ecr, measure\} & Basic set of native quantum operations \\
Quantum Volume (QV) & 128 & Characteristic circuit performance metric \\
Median $T_1$ & $\sim$263~µs & Energy relaxation time \\
Median $T_2$ & $\sim$177~µs & Dephasing time \\
Median single-qubit error (SX) & $2.4 \times 10^{-4}$ (0.024\%) & Average single-qubit gate error \\
Median two-qubit error (ECR) & $7.6 \times 10^{-3}$ (0.76\%) & Average two-qubit (ECR) gate error \\
Median readout error & $1.35 \times 10^{-2}$ (1.35\%) & Average qubit measurement error \\
Single-qubit gate duration (SX) & $\approx 60$~ns & Gate operation time \\
Two-qubit gate duration (ECR) & $\approx 660$~ns & Gate operation time \\
Readout duration & $\approx 700$~ns & Typical measurement time \\
Sampling period ($dt$) & $2.22 \times 10^{-9}$~s & Backend control electronics sampling step \\
Operating temperature & $\approx 15$~mK & Typical cryogenic environment \\
\hline
\end{tabular}
\\[4pt]
\footnotesize
All values represent median calibration data typical of Eagle-r3 (127-qubit) systems, such as \textit{ibm\_sherbrooke}.  
Actual values for the IBM--Cleveland Clinic System One may vary slightly due to routine recalibrations.
\end{table}

\newpage

\begin{figure}
    \centering
    \includegraphics[width=0.9\linewidth]{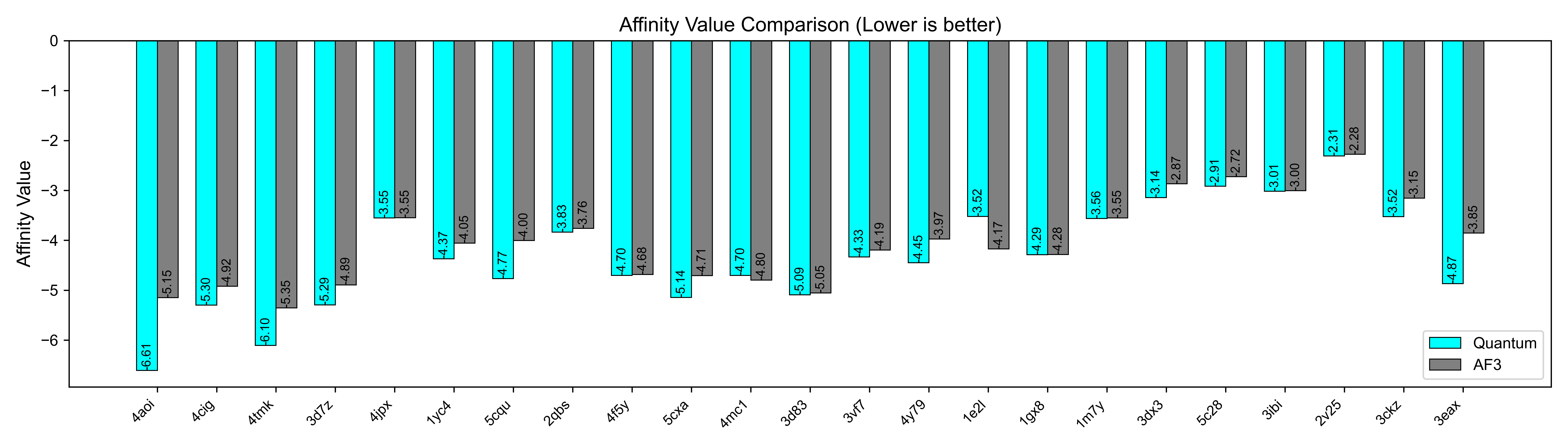}
    \caption{\textbf{RMSD comparison between quantum and AF3 predictions.}
    In 18 out of 23 test cases, quantum-predicted structures exhibit lower RMSD values relative to experimental structures compared to AF3, indicating improved geometric fidelity.}
    \label{fig:rmsd_bar}
\end{figure}

\begin{figure}
    \centering
    \includegraphics[width=0.9\linewidth]{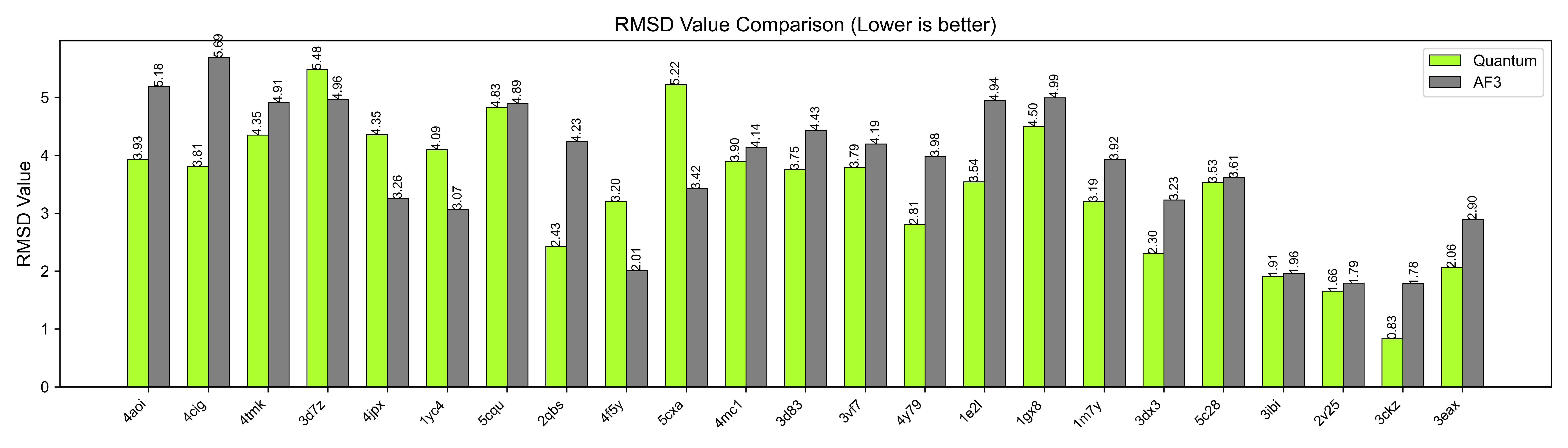}
    \caption{\textbf{Docking affinity scores of quantum vs. AF3-predicted structures.}
    Quantum structures achieved more favorable (lower) binding free energies in 21 of 23 cases, indicating better predicted protein-ligand interactions than AF3.}
    \label{fig:affinity_bar}
\end{figure}

\begin{figure}[t]
    \centering
    \includegraphics[width=0.9\linewidth]{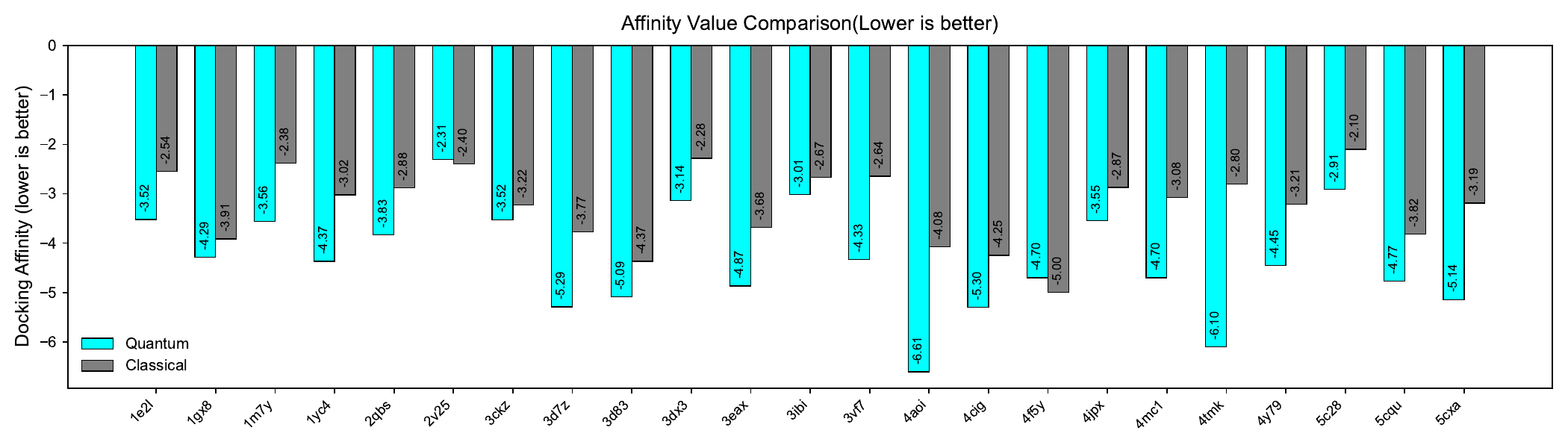}
    \caption{\textbf{Docking affinity scores of quantum vs. classical-predicted structures.}
    Quantum-predicted structures achieved more favorable (lower) binding free energies in 21 of 23 cases, indicating better predicted protein-ligand interactions and enhanced energetic realism compared to classical models.}
    \label{fig:affinity_bar_classical}
\end{figure}

\begin{figure}[t]
    \centering
    \includegraphics[width=0.9\linewidth]{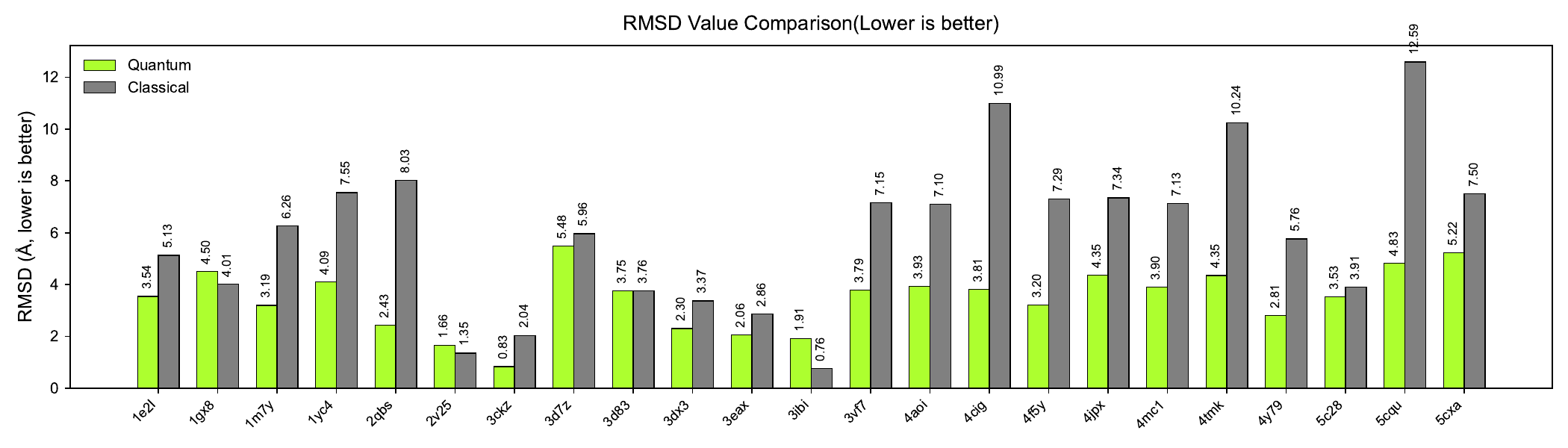}
    \caption{\textbf{RMSD comparison between quantum and classical predictions.}
    In 23 benchmark fragments, quantum-predicted structures exhibit lower RMSD values relative to experimental structures compared to classical models, indicating improved geometric fidelity and backbone stability, particularly in flexible or short-chain regions.}
    \label{fig:rmsd_bar_classical}
\end{figure}

\begin{figure*}
    \centering
    \includegraphics[width=0.8\textwidth]{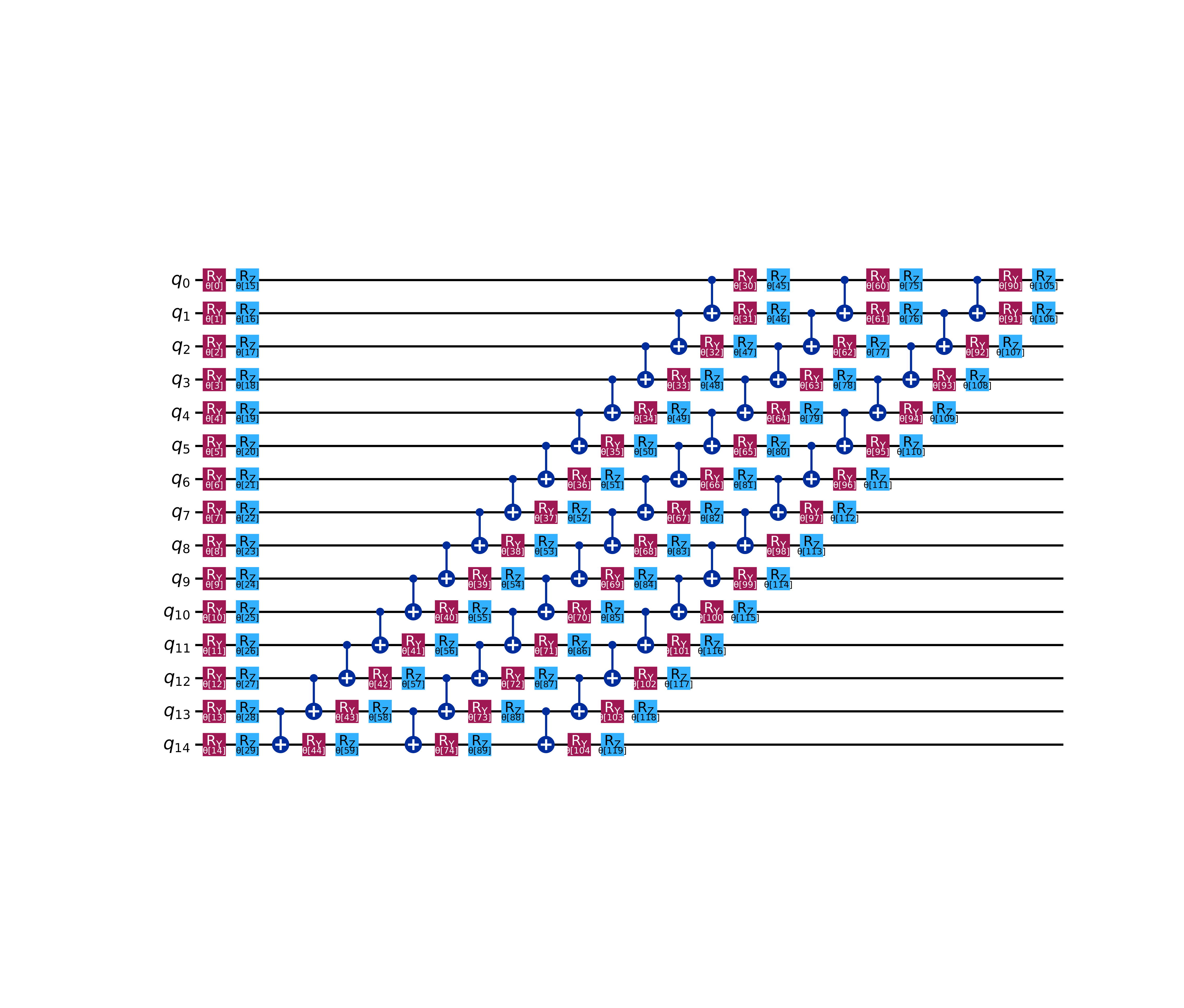}
    \caption{\textbf{Parameterized quantum circuit design used in VQE.}
    The ansatz follows the EfficientSU2 structure with alternating entanglement and rotation layers, enabling expressive quantum state preparation for structure optimization.}
    \label{fig:sup1}
\end{figure*}

\begin{figure}
    \centering
    \includegraphics[width=0.8\textwidth]{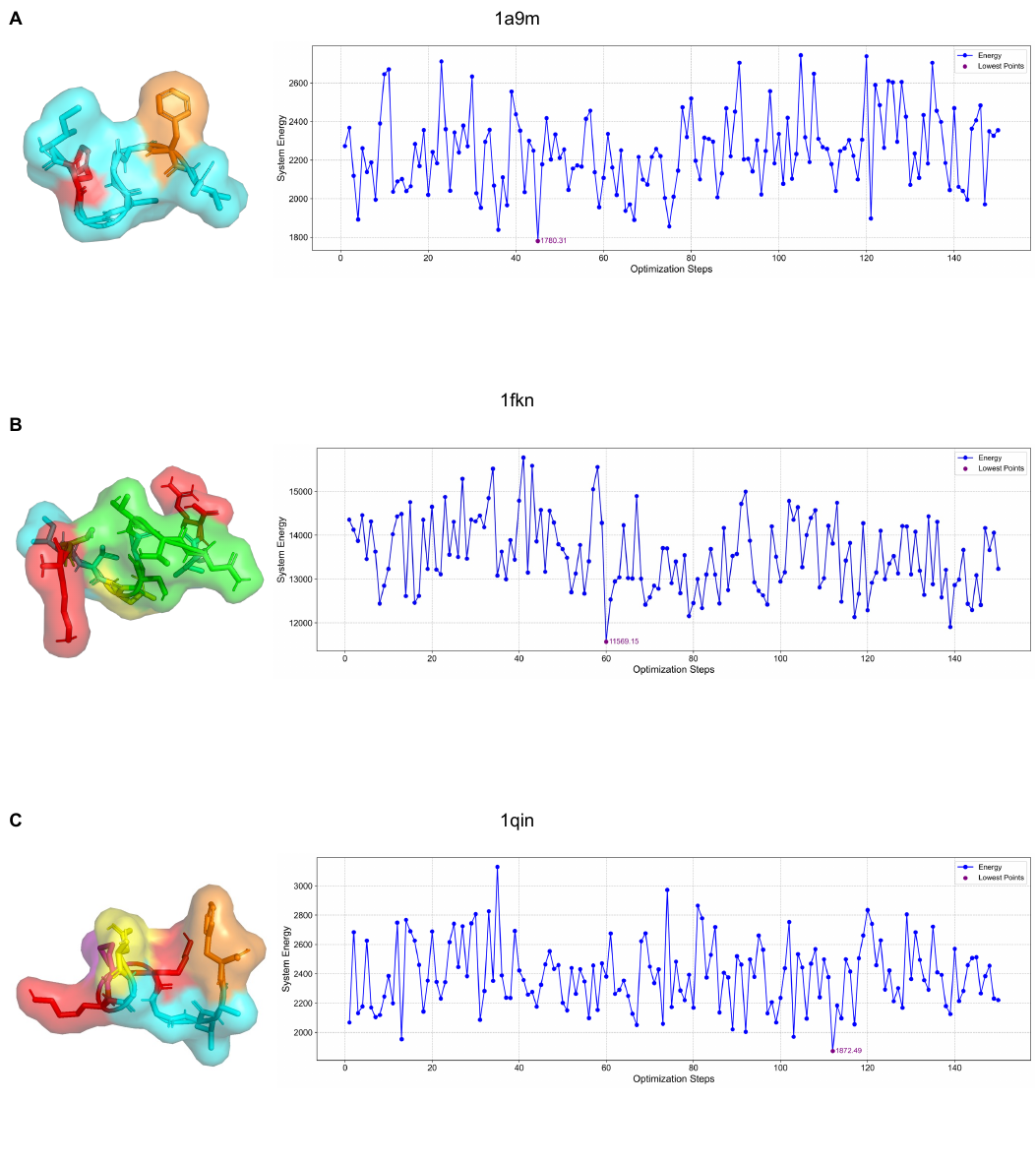}
    \caption{\textbf{Predicted structures and energy profiles for 1a9m, 1fkn, and 1qin.}
    Energy trajectories during VQE optimization are shown alongside the corresponding low-energy conformations for each fragment.}
    \label{fig:sup2}
\end{figure}

\begin{figure*}
    \centering
    \includegraphics[width=0.8\textwidth]{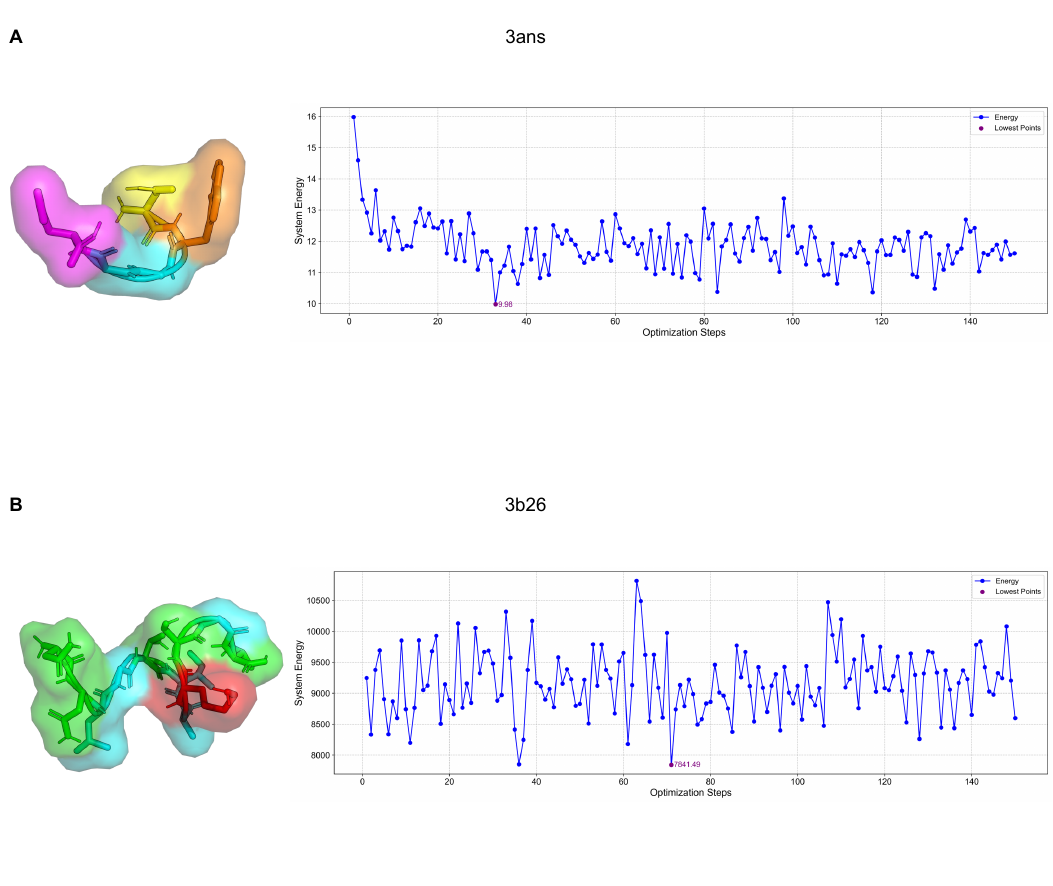}
    \caption{\textbf{Predicted structures and energy profiles for 3ans and 3b26.}
    Conformations sampled during VQE convergence illustrate how quantum-optimized structures correspond to energy minima.}
    \label{fig:sup3}
\end{figure*}

\begin{figure*}
    \centering
    \includegraphics[width=0.8\textwidth]{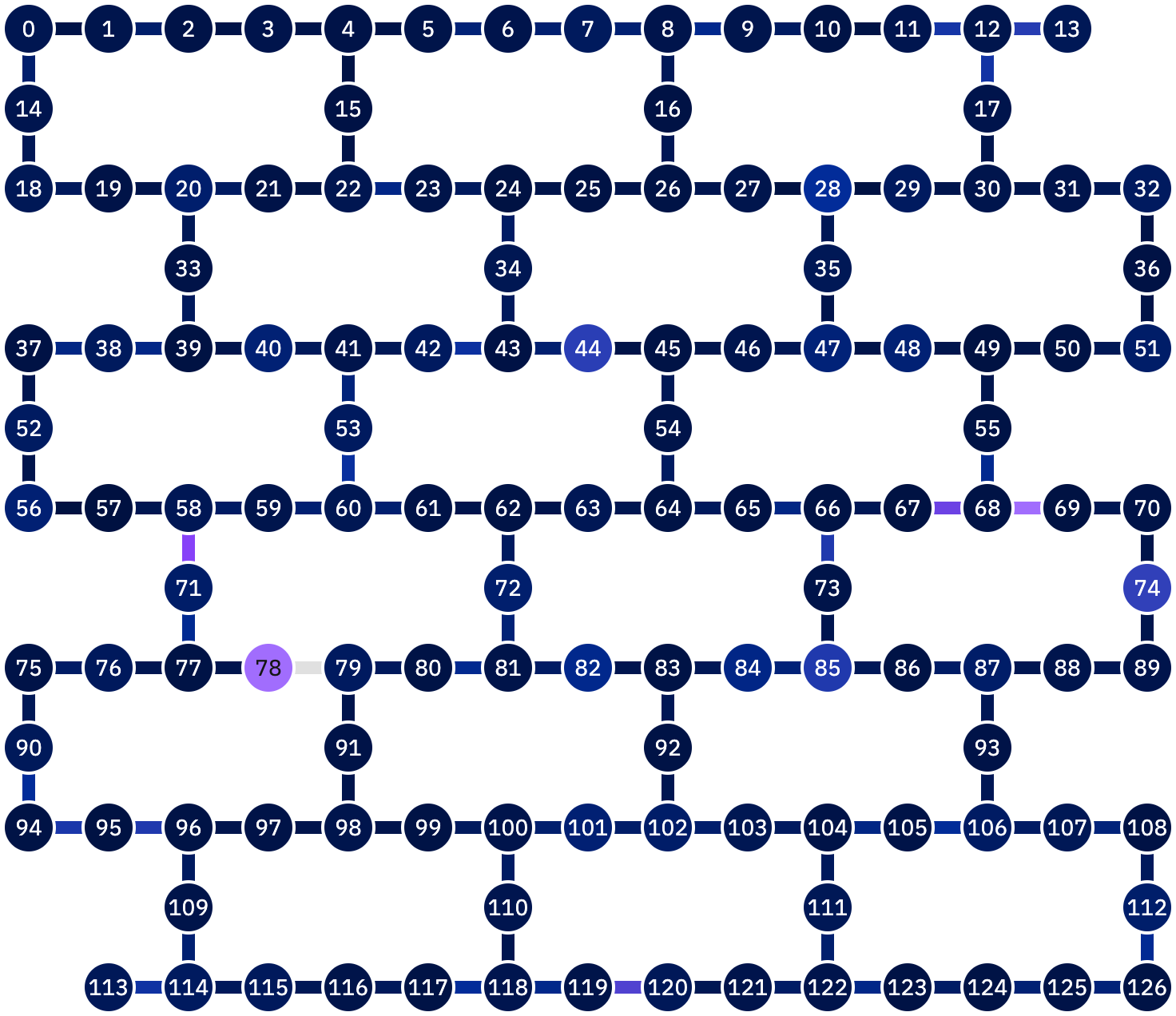}
    \caption{\textbf{Coupling map of the IBM-Cleveland Clinic quantum processor.}
    The hardware topology used in this study supports 127 qubits with directional two-qubit gate connectivity optimized for low error rates.}
    \label{fig:sup4}
\end{figure*}

\begin{figure*}
    \centering
    \includegraphics[width=0.9\textwidth]{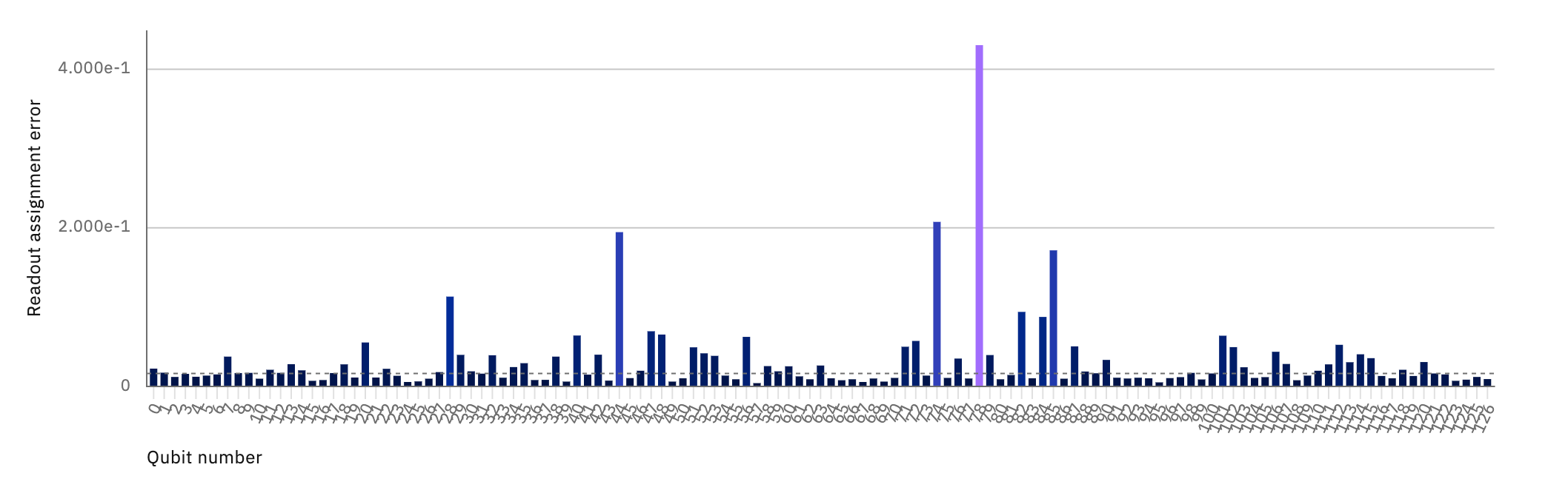}
    \caption{\textbf{Raw noise statistics distribution of the IBM-Cleveland Clinic quantum processor.}}
    \label{fig:sup5}
\end{figure*}

\begin{figure}[htbp]
    \centering
    \includegraphics[width=0.9\linewidth]{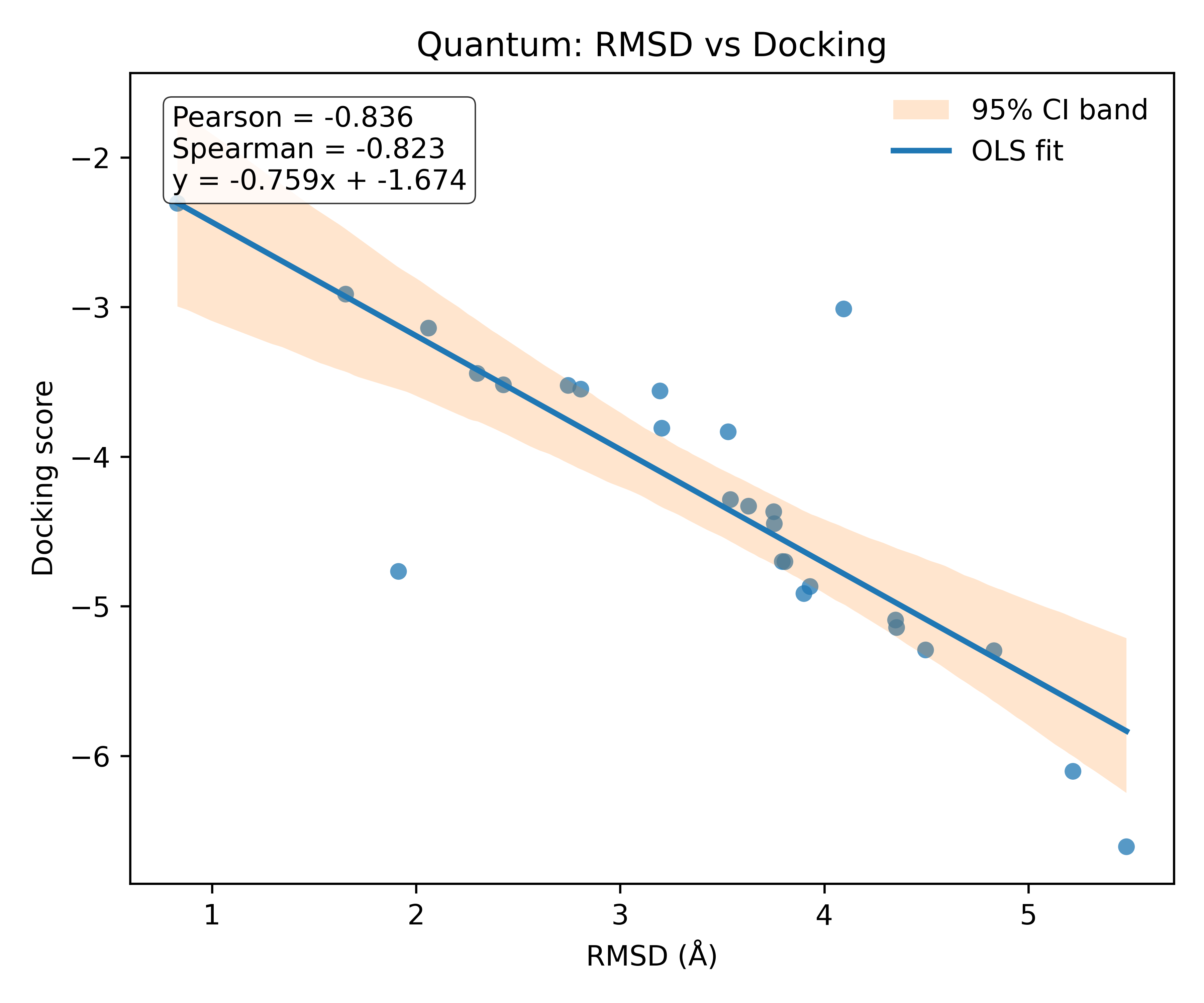}
    \caption{
        \textbf{Correlation analysis between RMSD and docking score for the quantum-predicted structures.}
        A strong negative correlation is observed, indicating that lower RMSD (i.e., higher structural accuracy) corresponds to stronger docking affinity. 
        The correlation coefficients (Pearson = $-0.836$, Spearman = $-0.823$) confirm a statistically significant relationship between geometric and energetic consistency. 
        Shaded areas represent the 95\% confidence intervals.
    }
    \label{fig:correlation_analysis}
\end{figure}





\end{document}